\documentclass[onecolumn,pra,amsmath,amssymb,nofootinbib,showkeys]{revtex4}

\usepackage{amsmath,amssymb,mathtools}
\usepackage{braket}
\usepackage{tikz}
\usetikzlibrary{quantikz,arrows.meta,decorations.pathmorphing}
\usepackage{float}
\usepackage{lipsum}
\usepackage{graphicx}
\usepackage{caption}

 \usepackage{tcolorbox} 

\begin{document}

\title{Quantum deliberating machines}

\author{Andrei Galiautdinov}
\email{ag1@uga.edu}
\affiliation{
Department of Physics and Astronomy, 
University of Georgia, Athens, Georgia 30602, USA
}

\date{\today}

\begin{abstract}
Within the familiar circuit-based quantum computational setting, 
we introduce and analyze a toy model of a quantum physical 
device capable of internal, self-referential deliberation. 
The key idea is to represent ``deliberation'' as a coherent 
iterative branching process, in which competing 
branch-dependent system evolutions are maintained in 
superposition, with additional control and memory registers 
recording branch histories, and the policy register adaptively 
biasing subsequent development. We provide explicit quantum 
circuit realizations, carry out detailed step-by-step derivations 
of the entangled control--memory--system--policy dynamics, 
and analyze the constraints imposed by the no-cloning theorem 
on coherent information flow. We carefully distinguish between 
internally adaptive and internally reinforced deliberations, 
proposing the architectures for both, and briefly discuss 
categorical and controlled--Stinespring reformulations, as well 
as their conceptual implications. The primary construction
models a memory-driven deliberation where the policy update 
depends on which actions were taken, not on their results. 
We also present a simple extension that allows for 
minimalistic, outcome-driven policy updates, implementing a 
coherent feedback loop that steers the system toward a 
target state regardless of initial branch-dependent evolution. 
This loop can be interpreted as a quantum autopilot or 
search-and-rescue mechanism, illustrating how a device 
can autonomously correct and optimize its internal strategy 
in superposition. Finally, we briefly consider various implementations 
of a dialogue that may take place between two deliberating machines. 
Taken together, this frames 
the proposed model as a plausible setting for exploring how 
such devices may maintain multiple alternatives in parallel, 
while performing an internal decision-making process 
through coherent branching, entanglement, adaptive policy 
updates, and policy-driven self-modifying unitary dynamics.
\end{abstract}


\maketitle


\section{Introduction}

Superposition is the central foundational concept of quantum 
mechanics \cite{Dirac1981, Peres1995}. In quantum computing 
a register may evolve under one of several unitaries depending 
on the state of a control qubit. Yet in the standard model of 
quantum computation, the system never evolves under a 
literal superposition of unitary operations. Instead, 
controlled unitaries are always block-diagonal, so the global 
dynamics is unitary while the subsystem evolves under one 
operation conditional on a control state \cite{NielsenChuang}. 
This structure underlies the concept of quantum parallelism 
which is exploited in algorithms such as Deutsch-Jozsa 
\cite{Deutsch1992} and Grover \cite{Grover1996}, where 
multiple classical alternatives are coherently processed, 
even though the underlying operations themselves are 
conditioned on the superposed inputs rather than themselves 
existing in a superposition. Consequently, standard quantum 
circuits operate with a fixed external control flow; they lack the 
self-referential architecture necessary for the system's own 
superposed history and internal policy to dynamically 
reconfigure the unitary operations acting upon the workspace.

This raises a conceptual question about what it would mean 
for a quantum device to \emph{simultaneously entertain 
competing evolutions} in an adaptive, iterative, self-referential 
sense. In classical cognitive science, deliberation is often 
modeled as the simultaneous consideration of multiple 
possible actions or thought streams before commitment 
to a particular outcome \cite{Kahneman2011, GoldShadlen2007, 
Miller2001, Posner1990, Petersen2012, Baddeley1974, Baddeley2000}. 
For example, perceptual decision-making is widely understood to 
involve parallel accumulation of evidence for different options, 
with the eventual choice depending on which process first reaches 
a threshold. More generally, working memory and attention 
mechanisms are thought to sustain alternative representations 
in parallel while cognitive control mechanisms weigh their relative 
importance \cite{Baars1988,Dehaene2001}.  
In artificial intelligence, similar 
principles are used in decision-theoretic settings, where multiple 
candidate hypotheses are maintained, tested, and 
updated in parallel, before one is actually selected (see, e.g., \cite{Sutton2018}). 
Such models embody parallelism at the algorithmic level, but are 
ultimately realized on deterministic or stochastic substrates that 
can only approximate true parallel exploration by probabilistic 
sampling. They cannot capture the full coherence inherent in 
quantum superposition, where alternatives coexist without 
collapsing until measurement.

Adopting this interpretive stance, we propose a self-referential
quantum computing architecture for a quantum analogue of such 
deliberative systems (which we call a \emph{quantum deliberating 
machine}, or QDM), consisting of control, memory, system, and 
policy registers, $(C,M,S,P)$, all quantum, that jointly evolve under global 
iterative unitary dynamics. Each iteration applies a branch-dependent 
system unitary, copies the control into a fresh memory register, and 
then updates the policy register conditionally on the recorded branch. 
At the same time, the system receives coherent feedback via 
controlled operations from the current state of $P$, so that the 
system's evolution is ``redirected'' by prior policy adaptation, before 
$P$ itself is further modified. Memory becomes entangled to track 
which computational branches were realized, while the policy 
register coherently adapts based on accumulated history. 
By iterating this process, the QDM naturally implements a form of 
internally coherent decision-making in which multiple hypothetical 
evolutions are entertained and weighted before an effective choice 
finally emerges (cf.\ alternative proposal of Ref.\  \cite{Dunjko2015}).

Conceptually analogous to quantum parallelism in Deutsch-Jozsa 
and Grover (also \cite{Shor1997}), the QDM extends the parallelism 
idea by coherent superposition of entire branch-dependent unitary 
evolutions while entangling system, memory, and policy registers 
across multiple iterations. This architectural structure implements 
self-modifying global evolution in which branch histories are tracked 
and preferences are coherently updated, providing a formal model 
for internally simulating alternative dynamical pathways ({\it cf}.\ 
\cite{Adlam2025}). We formulate the QDM primarily within the 
pure-state, unitary picture, since this provides a direct and 
transparent account of its iterative adaptive deliberative 
dynamics. Reduced states do appear in our discussion 
of the classical appearance of memory marginals and their relation 
to the no-cloning theorem, but these are still viewed as byproducts 
of an underlying pure-state evolution. 

While the present work draws heuristic inspiration from quantum 
computing algorithms, the analogy is conceptual rather than formal. 
We stress that, whereas Deutsch-Jozsa and Grover exploit parallel 
evaluation of multiple input states under a fixed unitary, the QDM 
places entire alternative branching evolutions themselves in coherent 
superposition, allowing the dynamics of the system, memory, and 
policy registers to simultaneously evolve along multiple trajectories. 

Our proposal of a QDM builds on a diverse body of research in 
quantum computation, higher-order quantum processes, quantum 
learning, and the philosophy of quantum information in general. 
Feynman’s seminal works first established the motivation for quantum 
computation by arguing that classical devices face exponential 
obstacles in simulating quantum systems, while inherently quantum 
simulators could overcome this bottleneck 
\cite{Feynman1982,Feynman1986} (also \cite{Lloyd1996}). 
Nielsen and Chuang’s standard reference text \cite{NielsenChuang} 
formalized the quantum circuit model by emphasizing unitary dynamics 
and measurement as the primitive resources of quantum information 
processing. Our proposal is directly inspired by these foundations, 
which it extends by considering coherent superpositions not merely 
of states but of entire branching unitary evolutions, thereby treating 
the machine’s ``deliberations'' as quantum processes in their own right.

An important precedent for this conceptual move was the development 
of higher-order quantum maps, beginning with the theory of 
supermaps introduced by Chiribella, D’Ariano, and Perinotti 
\cite{Chiribella2008} and further elaborated in the quantum 
networks setting \cite{Chiribella2009}. These works showed 
that quantum operations themselves can be treated as inputs 
to coherent transformations. They established that the formalism of 
quantum mechanics can accommodate such process-level superpositions. 
Notable examples include the quantum switch, where channels can be 
applied in an indefinite causal order, and the demonstration that quantum 
correlations can exist without definite causal structure 
\cite{Chiribella2013, Oreshkov2012}. Our QDM is conceptually aligned with 
this line of research, but emphasizes a distinct interpretation. Instead of 
superposing orders of operations, it forms superpositions of alternative 
dynamical evolutions mediated by additional memory and policy 
registers that guide the internal deliberation, a perspective that may also 
admit categorical reformulation (cf.~\cite{Abramsky2004}).

Connection to quantum learning further sharpens this perspective. 
Dunjko, Taylor, and Briegel introduced quantum reinforcement learning 
agents with internal policy registers \cite{Dunjko2016}, showing that 
quantum systems can provide advantages in adaptive decision-making 
scenarios. Our model extends this idea by treating the policy register 
not as a fixed parameter, but as a coherently evolving degree of freedom 
entangled with memory, thereby embedding deliberative dynamics 
directly into the quantum substrate.\footnote{In \cite{Dunjko2016}, 
the agent’s policy register interacts with both internal and 
environment-driven inputs, effectively incorporating external signals 
into its updates. By contrast, our QDM model is formulated entirely 
in pure-state form; the policy register coherently evolves as an internal 
degree of freedom entangled with memory, without requiring an external 
environment.} This property of the model is constrained by the 
no-cloning theorem \cite{Wootters1982,Barnum1996}, which prohibits 
unrestricted copying of quantum states. Instead of being a limitation, 
this constraint motivates an entanglement-based memory structure. 
The QDM records its branching deliberations not by duplicating states, 
but by encoding them in correlated registers, much as quantum 
mechanics naturally prescribes.

Taken together, these strands provide a coherent backdrop for our 
construction. Feynman’s vision motivates the physical grounding of 
quantum computation, Nielsen and Chuang supply the standard model 
for us to use, Chiribella and collaborators formalize superpositions at 
the level of processes, Oreshkov and colleagues show that causal structure 
itself admits quantum generalization, Dunjko et al.\ highlight the role of 
quantum policy registers in adaptive learning, and Wootters and Zurek 
guarantee that memory must be entangled rather than cloned. 
What distinguishes the QDM is not any single ingredient, but the 
synthesis --- a machine whose internal, self-reflective 
deliberations are modeled as coherent superpositions of alternative 
unitary pathways, constrained by quantum principles and guided by 
adaptive quantum policies (cf.~\cite{Bera2017, Bera2019, Rubino2017, 
Rubino2021, Elliott2022, Stromberg2024, Sultanow2025}).

\section{The Basic Elements of the QDM}

\subsection{Superposition of Unitary Evolutions: Basic Construction}

\begin{figure}[H]
\centering
\begin{quantikz}
\lstick{$C$} & \gate{\alpha\ket{0}+\beta\ket{1}}  & \ctrl{1} & \qw \\
\lstick{$S$} & \qw & \gate{U_0/U_1} & \qw
\end{quantikz}
\caption{Controlled-unitary evolution: $C$ selects between 
$U_0$ or $U_1$ on $S$, creating a coherent superposition 
of process paths.}
\label{fig:CU-diagram}
\end{figure}
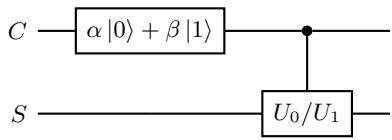

Consider a system $S$ (\emph{workspace}), such as, e.g., a single qubit, 
and two Hamiltonians, $H_0$ and $H_1$, generating distinct unitary 
evolutions,
\begin{equation}
U_0 = e^{-i H_0 t}, \qquad U_1 = e^{-i H_1 t}.
\end{equation}
We introduce a control qubit $C$ (\emph{attention pointer}, 
or \emph{attentional control}), such that the evolution of the 
combined system $(C,S)$ obeys,
\begin{align}
\ket{0}_{C}\ket{\psi}_{S} &\mapsto \ket{0}_{C} (U_{0}\ket{\psi}_{S}), \\
\ket{1}_{C}\ket{\psi}_{S} &\mapsto \ket{1}_{C} (U_{1}\ket{\psi}_{S}),
\end{align}
which corresponds to applying a controlled-unitary 
operator $U$ with $C$ as control
(\cite{Aharonov1990}, \cite{Oi2003}; cf.~\cite{Bera2019}; also Appendix \ref{sec:stinespring_pure_state}),
\begin{equation}
U = \ket{0}\bra{0}_C \otimes U_0 + \ket{1}\bra{1}_C \otimes U_1 .
\end{equation}
If control $C$ is initially prepared in a normalized superposition,
\begin{equation}
\ket{\phi}_C = \alpha \ket{0}_C + \beta \ket{1}_C, \quad |\alpha|^2 + |\beta|^2 = 1,
\end{equation}
then the joint state after application is
\begin{equation}
\ket{\Psi}_{CS} = \alpha \ket{0}_C \otimes U_0 \ket{\psi}_S 
+ \beta \ket{1}_C \otimes U_1 \ket{\psi}_S.
\end{equation}
Intuitively, this joint pure state represents a superposition of two 
alternative competing evolutions, entangled with the control 
qubit, which acts as a ``quantum switch'' \cite{Bera2019}, thus 
enabling the machine to coherently maintain both candidate 
processes in parallel, Fig.\ \ref{fig:CU-diagram}.
At some later time, the control qubit can be measured, selecting 
one branch, or it can be retained as part of a larger entangled 
register. 

The above minimal model can be enriched. A memory register 
$M$ can coherently record the labels of actions taken 
(and, possibly, the outcome of the evaluation) by 
a controlled-NOT from $C$ to $M$. A policy register $P$ may 
store amplitudes that are coherently updated by a unitary 
controlled by $M$, representing a reversible quantum learning 
step. The entire $(C,M,S,P)$ system then unitarily evolves, with 
entanglement carrying the record of past actions (and outcomes) 
and updating future tendencies.

This extension suggests an architecture for quantum agents: 
rather than classically re-weighting probabilities after each 
outcome, the agent maintains coherent policy superpositions 
that adapt through reversible entanglement with memory.

\subsection{Internal Memory Extension}

\subsubsection{Branch Recording}

\begin{figure}[H]
\centering
\begin{quantikz}
\lstick{$C$} & \gate{\alpha\ket{0}+\beta\ket{1}}  &\ctrl{2} & \ctrl{1} & \qw \\
\lstick{$M$} & \qw & \qw &\targ{} & \qw \\
\lstick{$S$} & \qw & \gate{U_0/U_1} & \qw &\qw
\end{quantikz}
\caption{Memory recording: $M$ entangled with $C$ to 
record the branch without measurement.}
\label{fig:CMNOT-diagram}
\end{figure}
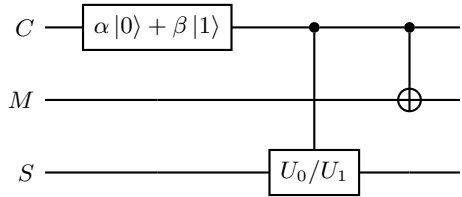

To model the QDM's ``internal record'' (or ``internal diary'') of 
which process was 
active, we add a \emph{(working) memory} register (qubit), $M$, 
initialized in $\ket{0}_M$. A simple mechanism is to 
coherently copy the control's value,
\begin{equation}
\label{eq:copy_control's_value}
\ket{c}_{C}\ket{0}_{M} \mapsto \ket{c}_{C}\ket{c}_{M},
\end{equation}
where $c\in\{0,1\}$ labels the computational basis states,
or,
\begin{align}
\ket{0}_{C}\ket{0}_{M} &\mapsto \ket{0}_{C}\ket{0}_{M}, \\
\ket{1}_{C}\ket{0}_{M} &\mapsto \ket{1}_{C}\ket{1}_{M}.
\end{align}
After the $C$-controlled evolution of $S$, we apply a CNOT 
gate with $C$ as control and $M$ as target, and get,
\begin{equation}
\ket{\Psi}_{CMS}
= \alpha \ket{0}_C \ket{0}_M \otimes U_0 \ket{\psi}_S
+ \beta \ket{1}_C \ket{1}_M \otimes U_1 \ket{\psi}_S .
\end{equation}
This entangles the memory qubit with the control qubit, 
as depicted in Fig.\ \ref{fig:CMNOT-diagram}. 

The memory now coherently correlates with which process 
was applied (effectively labeling the computational branch), 
without requiring measurement (and, thus, without collapsing 
the superposition). It can be viewed as an internal ``scratchpad'' 
for the QDM to track its evolution. This setup enables coherent 
feedback mechanisms to conditionally act on the system based 
on recorded branch information (see below).

\subsubsection{No-cloning Considerations}

As is well known, the CNOT operation creates an entangled state 
rather than a product state, consistent with the no-cloning theorem. 
If the control is in a superposition
\begin{equation}
\ket{\phi}_C = \alpha \ket{0}_C + \beta \ket{1}_C ,
\end{equation}
then after the CNOT with $M$ initially in $\ket{0}_M$ we obtain
\begin{equation}
\ket{\Psi}_{CM} = \alpha \ket{0}_C \ket{0}_M + \beta \ket{1}_C \ket{1}_M.
\end{equation}
This state \emph{not} equal to
\begin{equation}
(\alpha \ket{0} + \beta \ket{1})_C \otimes (\alpha \ket{0} + \beta \ket{1})_M,
\end{equation}
which would be required for cloning. The reduced state of the memory is
\begin{equation}
\rho_M = \mathrm{Tr}_C \left[ \ket{\Psi}_{CM}\bra{\Psi} \right]
= |\alpha|^2 \ket{0}\bra{0}_M + |\beta|^2 \ket{1}\bra{1}_M,
\end{equation}
a mixed state diagonal in the computational basis. Thus the memory 
only acquires classical information about the branch label, 
not a full copy of the quantum amplitudes (see Sec.\ \ref{sec:marginals}, 
and especially Sec.\ \ref{sec:no-cloning-detailed}, for more careful 
discussion). Copying of arbitrary quantum states remains forbidden, 
so the no-cloning theorem is not violated. This corresponds to the 
idea of \emph{broadcasting} classical information while preserving 
quantum coherence \cite{Barnum1996}.

\subsection{Quantum Policy Register}

\subsubsection{Conceptual Overview}

To simulate adaptive behavior, we add a policy register $P$ 
(\emph{internal model, belief state, or strategy representation}), which 
encodes the machine's evolving internal ``decision policy'' 
(cf.~\cite{Sutton2018}). This policy register $P$ is a qubit whose 
amplitudes encode the machine’s internal preference for selecting 
each branch. It is initialized in
\begin{equation}
\ket{\pi}_P = \gamma \ket{0}_P + \delta \ket{1}_P.
\end{equation}
This quantum register evolves conditionally on the memory register, 
allowing the machine to coherently adapt its internal preferences. 
The policy register enables a reversible, internal form of learning. 
Unlike classical probabilistic choices, the amplitudes in $P$ remain 
fully coherent with the rest of the system. This mechanism provides 
a conceptual model for internal adaptive decision-making.

\subsubsection{Conditional Policy Update}

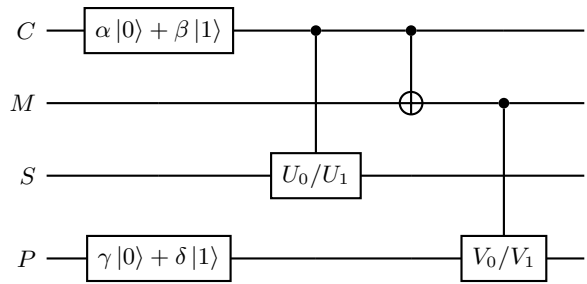
\begin{figure}[H]
\centering
\begin{quantikz}
\lstick{$C$} & \gate{\alpha\ket{0}+\beta\ket{1}}     &\ctrl{2}                & \ctrl{1} & \qw     &\qw\\
\lstick{$M$} & \qw                                                   & \qw                   &\targ{}  & \ctrl{2} & \qw \\
\lstick{$S$} & \qw                                                     & \gate{U_0/U_1} & \qw     &\qw       &\qw \\
\lstick{$P$} & \gate{\gamma\ket{0}+\delta \ket{1}} & \qw                   & \qw     & \gate{V_0/V_1} & \qw
\end{quantikz}
\caption{Quantum policy update, $V_{\rm update}$: $P$ is updated 
conditionally on memory $M$.}
\label{fig:P-update-diagram}
\end{figure}

Define a controlled-unitary on $P$ using $M$ as control,
\begin{equation}
V_{\rm update} = \ket{0}\bra{0}_M \otimes V_0 + \ket{1}\bra{1}_M \otimes V_1,
\end{equation}
where $V_0$ and $V_1$ as small rotations on $P$. Applying $V_{\rm update}$ yields
\begin{equation}
\ket{\Psi}_{CMSP} = \alpha \ket{0}_C \ket{0}_M \otimes 
U_0 \ket{\psi}_S \otimes V_0 \ket{\pi}_P + \beta \ket{1}_C \ket{1}_M 
\otimes U_1 \ket{\psi}_S \otimes V_1 \ket{\pi}_P.
\end{equation}
The policy register becomes entangled with the memory, allowing 
conditional coherent updates (Fig.\ \ref{fig:P-update-diagram}). 
No measurement occurs, maintaining a pure-state description. 
This step models a quantum analogue of adaptive policy update 
(reminiscent of reinforcement learning, but without evaluative reward 
signals).

\subsection{Policy-controlled feedback}

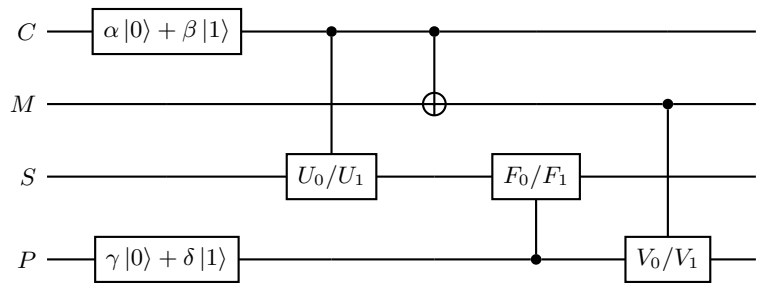
\begin{figure}[H]
\centering
\begin{quantikz}[row sep=0.5cm, column sep=0.6cm]
\lstick{$C$}            & \gate{\alpha\ket{0}+\beta\ket{1}} & \ctrl{2}                & \ctrl{1}    &\qw                                              & \qw                                 & \qw \\
\lstick{$M$}  & \qw & \qw                                  & \targ{}              &\qw                         & \ctrl{2}                             & \qw \\
\lstick{$S$}            &\qw  & \gate[wires=1]{U_0/U_1}  &\qw  &   \gate[wires=1]{F_0/F_1}    & \qw                                 & \qw \\
\lstick{$P$}            & \gate{\gamma\ket{0}+\delta \ket{1}} & \qw            &\qw                  &   \ctrl{-1}                                & \gate[wires=1]{V_0/V_1}  & \qw
\end{quantikz}
\caption{Policy-controlled feedback $F_{\rm feedback}$ 
(performed before policy update): $S$ undergoes additional 
unitary evolution, conditioned on policy $P$.}
\label{fig:P-modulated-feedback-diagram}
\end{figure}

Finally, let us suppose that the policy acts as an additional conditional 
unitary (feedback) $F_0$ or $F_1$ on the system qubit $S$, depending 
on whether $P$ is in the state $\ket{0}$ or $\ket{1}$. The corresponding 
controlled-unitary on $S$ controlled by $P$ is given by
\begin{equation}
F_{\rm feedback} = \ket{0}\bra{0}_P \otimes F_0 + \ket{1}\bra{1}_P \otimes F_1,
\end{equation}
which (and this is conceptually important) should be applied {\it before} 
the policy update.
This expression makes clear that the system $S$ now evolves under 
a composition of both its original branch-dependent dynamics 
($U_0$ or $U_1$) and the learned influence of the policy register 
($F_0$ or $F_1$), Fig.\ \ref{fig:P-modulated-feedback-diagram}.

\subsection{On the role of the policy register: $P \to S$ versus $P \to C$}

In constructing the QDM, one subtle but important design choice 
concerns how the policy register $P$ interacts with the rest of 
the global system. Two natural architectures may be envisioned:

\begin{enumerate}
    \item[(i)] \textbf{Direct policy control of the system ($P \to S$):} 
    the policy register directly determines which unitary evolution 
    the system qubit $S$ undergoes at each iteration. Here the meaning 
    is transparent: the internal policy represents adaptive rules or 
    preferences, and these act immediately to shape the system's 
    evolution. This is the architecture adopted in our main construction.
    It is used here to implement what we term \emph{adaptive} 
    deliberation.

    \item[(ii)] \textbf{Indirect policy control via the control register 
    ($P \to C \to S$):} the policy register instead controls the control 
    qubit $C$, which then selects which unitary is applied to $S$. 
    In this model, $P$ does not directly manipulate $S$, but rather 
    ``programs'' the control channel that governs $S$'s possible 
    dynamics. This structure is somewhat more circuit-like, since it 
    preserves the canonical role of $C$ as the selector of alternatives.
    It may be used to implement the \emph{reinforced} deliberation
    (Section \ref{sec:reinforcement}).
\end{enumerate}

Both designs preserve the essential feature at the center of the QDM 
proposal---the simultaneous maintenance of \emph{parallel processes}, 
corresponding to the coherent superposition of distinct dynamical 
evolutions. In either case, the control register $C$ continues to serve 
as the quantum branch marker, entangling with the system $S$ and the 
memory register $M$, so that the machine maintains awareness of multiple 
alternatives in parallel. The policy register $P$, whether acting 
directly on $S$ or indirectly through $C$, provides the adaptive 
component. It modifies how the two alternatives are entertained, given 
the information encoded in memory.

We chose the $P \to S$ architecture as our preferred formulation 
mainly for the following two reasons. First, it eliminates unnecessary 
indirection; the policy acts where it matters most, namely on the system's 
actual evolution. Second, it matches the intuition from reinforcement 
learning, where policies determine actions rather than controllers of actions. 

Nevertheless, the alternative $P \to C \to S$ architecture remains 
valuable as a conceptual comparison (see Section \ref{sec:reinforcement}
for its potential use to model reinforcement in decision making). 
It highlights that the QDM is not tied to a single implementation 
choice, but rather to the deeper idea that coherent superpositions 
of evolutions can be guided by adaptive quantum registers. 
This flexibility suggests that more elaborate hybrid architectures 
could be constructed, where policy acts at multiple levels 
(both on $C$ and on $S$), thereby enriching the repertoire of internal 
adaptive responses available to the machine.

\subsection{Evaluation Step}

At some chosen time, the machine performs an ``evaluation'' by 
measuring the control register $C$ in the computational basis. The 
outcome selects one branch, collapsing the superposition into either 
the $U_{0}$- or the $U_{1}$-generated component. This models the act of 
committing to a decision or action after maintaining multiple alternatives 
in parallel. Since each iteration introduces a fresh memory register, 
the identities of branches evolve dynamically (Sec.\ \ref{sec:branches}); 
what persists is not the branch itself but the structure of alternatives 
that the QDM entertains. (In architectures with $P\to C$ 
feedback, the evaluation would select among adaptively reshaped 
branch amplitudes; see Sec.~\ref{sec:reinforcement}).

\subsection{Pure-State Formulation}

The key observation is that the entire description can be maintained 
in pure-state form. The composite state of the QDM is always pure; 
density matrices are not required unless one wishes to trace out 
subsystems or introduce open-system noise. The pure-state perspective 
makes transparent the notion that all alternative branches coexist in 
superposition, with their distinct histories recorded in the memory registers.

\section{Summary of the Model}
\label{sec:summary-of-the-model}

The operational definition of our QDM toy model can now be summarized 
as follows (see Fig.\ \ref{fig:QDM-overview-diagram}):

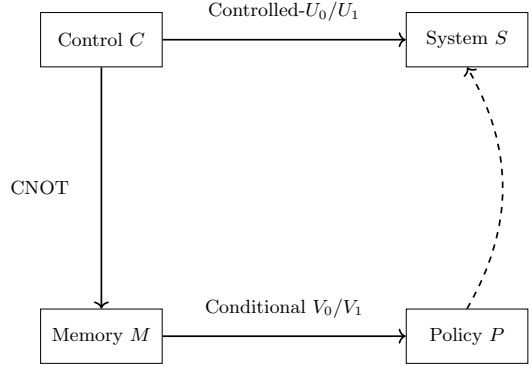
\begin{figure}[h]
\centering
\scalebox{0.8}{\begin{tikzpicture}[node distance=4.0cm, every node/.style={draw, rectangle, minimum width=2cm, minimum height=0.9cm, align=center}]
\node (C) {Control $C$};
\node[right=of C] (S) {System $S$};
\node[below=of C] (M) {Memory $M$};
\node[below=of S] (P) {Policy $P$};

\draw[->, thick] (C) -- (S) node[midway, above, draw=none] {Controlled-$U_0/U_1$};
\draw[->, thick] (C) -- (M) node[midway, left, draw=none] {CNOT};
\draw[->, thick] (M) -- (P) node[midway, above, draw=none] {Conditional $V_0/V_1$};
\draw[->, dashed, thick, bend right] (P.north) to (S.south);

\end{tikzpicture}
}
\caption{Overview of the QDM. Control qubit $C$ determines the 
superposed unitary applied to system $S$, memory $M$ records 
branch information, and policy $P$ adapts conditionally. 
Dashed arrow indicates possible feedback in iterative evolution.}
\label{fig:QDM-overview-diagram}
\end{figure}
 
Denote by $C$ the control qubit initialized in
\begin{equation}
\ket{\phi}_{C}^{(0)}=\alpha \ket{0}_{C} + \beta \ket{1}_{C}, \quad |\alpha|^2 + |\beta|^2 = 1.
\end{equation}
Also denote by $S$ the system qubit initialized in $\ket{\psi}_S^{(0)}$; 
by $M^{(k)}$ the memory qubit written in iteration $k$ and initialized 
in  $\ket{0}_{M^{(k)}}$; and by $P$ the policy qubit initialized in 
\begin{equation}
\ket{\pi}_P^{(0)} = \gamma \ket{0}_P + \delta \ket{1}_P.
\end{equation} 
Then, the global initial state of the machine is
\begin{align}
\ket{\Psi}_{\rm uncorr}^{(0)}
&=
\ket{\phi}_{C}^{(0)}\ket{0}_{M^{(1)}}\ket{\psi}_S^{(0)}\ket{\pi}_P^{(0)}
\nonumber \\
\label{eq:uncorrelated-initialization}
&=
\Big(
\alpha\gamma\ket{0}_{C}\ket{0}_{P}
+
\alpha\delta\ket{0}_{C}\ket{1}_{P}
+
\beta\gamma\ket{1}_{C}\ket{0}_{P}
+
\beta\delta\ket{1}_{C}\ket{1}_{P}
\Big)
\ket{0}_{M^{(1)}}\ket{\psi}_S^{(0)}
\qquad {\rm (uncorrelated\; initialization)},
\end{align}
with $C$ and $P$ initially uncorrelated.

Alternatively, we may initialize the system with $C$ and $P$ correlated, 
by acting on $\ket{\Psi}_{\rm uncorr}^{(0)}$ with either $C \to P$ 
or $P \to C$ controlled-NOT. In this work, we adhere to the $C \to P$ 
convention, in which case the initial state takes the ``correlated'' form,
\begin{align}
\ket{\Psi}_{\rm corr}^{(0)}
&=
\Big(
\ket{0}\bra{0}_C \otimes I_P + \ket{1}\bra{1}_C \otimes X_P
\Big)
\ket{\Psi}_{\rm uncorr}^{(0)}
\nonumber \\
\label{eq:correlated-initialization}
&=
\Big(
\alpha\gamma\ket{0}_{C}\ket{0}_{P}
+
\alpha\delta\ket{0}_{C}\ket{1}_{P}
+
\beta\delta\ket{1}_{C}\ket{0}_{P}
+
\beta\gamma\ket{1}_{C}\ket{1}_{P}
\Big)
\ket{0}_{M^{(1)}}\ket{\psi}_S^{(0)}
\qquad {\rm (correlated\; initialization)}.
\end{align}

Let the branch labels be $b\in\{0,1\}$ and the policy component labels 
$p\in\{0,1\}$. For iteration $k$ define:
\begin{itemize}
  \item branch-dependent system unitaries $U^{(k)}_b$ (applied to $S$ when $C$ has value $b$),
  \item policy-controlled feedback unitaries $F^{(k)}_p$ (applied to $S$ conditioned on the current state of $P$),
  \item branch-dependent policy updates $V^{(k)}_b$ (applied to $P$ conditioned on the recorded branch $b$).
\end{itemize}
Write the controlled-$U$ on $C\otimes S$ for iteration $k$ as
\[
  \mathcal{U}^{(k)} = \ket{0}\bra{0}_C \otimes U^{(k)}_0 + \ket{1}\bra{1}_C \otimes U^{(k)}_1,
\]
the controlled-$F$ on $P\otimes S$ as
\[
  \mathcal{F}^{(k)} = \ket{0}\bra{0}_P \otimes F^{(k)}_0 + \ket{1}\bra{1}_P \otimes F^{(k)}_1,
\]
and the controlled-$V$ on $M^{(k)}\otimes P$ as
\[
  \mathcal{V}^{(k)} = \ket{0}\bra{0}_{M^{(k)}} \otimes V^{(k)}_0 + \ket{1}\bra{1}_{M^{(k)}} \otimes V^{(k)}_1.
\]
Let $\mathrm{CNOT}_{C\to M^{(k)}}$ denote the CNOT gate with control $C$ and target $M^{(k)}$.

Then the \emph{single iteration $k$ update map} is defined as
\begin{equation}
  \label{eq:QDM_iteration_map}
 \mathcal{W}^{(k)} = \mathcal{V}^{(k)} \,  \mathcal{F}^{(k)} \, 
 \mathrm{CNOT}_{C\to M^{(k)}} \, \mathcal{U}^{(k)},
\end{equation}
where the policy-controlled feedback $ \mathcal{F}^{(k)}$ to $S$ is 
applied just before the corresponding policy update $\mathcal{V}^{(k)}$ to $P$. 
This ordering ensures that feedback uses the \emph{current state of 
the policy register} $P$ before it is updated by $M$ in this iteration, 
maintaining a coherent flow of adaptive influence on $S$. 
Eq.\ (\ref{eq:QDM_iteration_map}) represents the unitary operator on the joint 
space of all registers involved in step $k$.

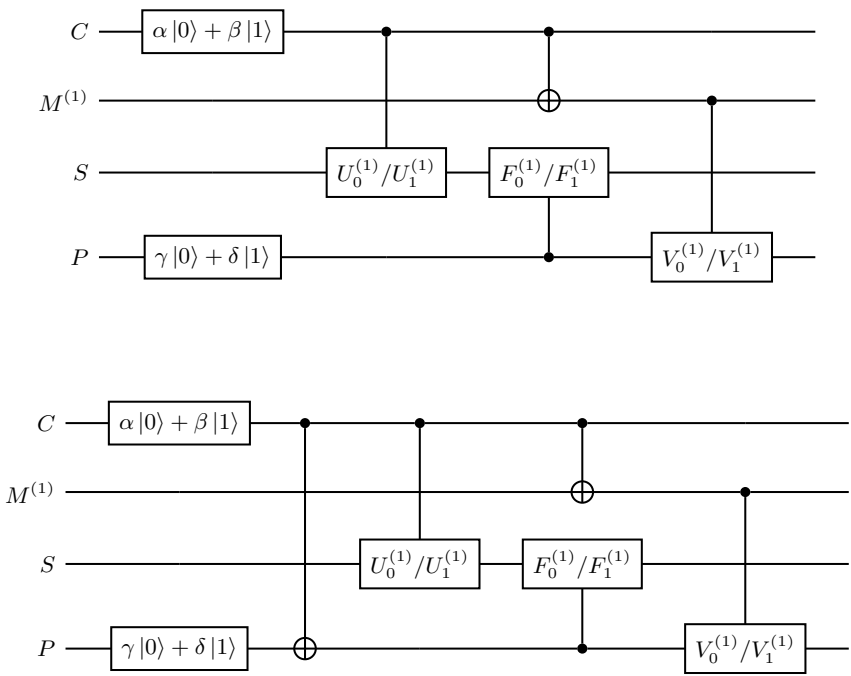
\begin{figure}[H]
\centering
\scalebox{0.95}{\begin{quantikz}[row sep=0.5cm, column sep=0.6cm]
\lstick{$C$}            & \gate{\alpha\ket{0}+\beta\ket{1}} & \ctrl{2}                              & \ctrl{1}                                        & \qw                                 & \qw \\
\lstick{$M^{(1)}$}  & \qw & \qw                                  & \targ{}                                        & \ctrl{2}                             & \qw \\
\lstick{$S$}            &\qw  & \gate[wires=1]{U_0^{(1)}/U_1^{(1)}}   &   \gate[wires=1]{F_0^{(1)}/F_1^{(1)}}    & \qw                                 & \qw \\
\lstick{$P$}            & \gate{\gamma\ket{0}+\delta \ket{1}}& \qw                                    &   \ctrl{-1}                                & \gate[wires=1]{V_0^{(1)}/V_1^{(1)}}  & \qw
\end{quantikz}
}
\vskip10pt
\vspace{1cm} 
\scalebox{0.95}{\begin{quantikz}[row sep=0.5cm, column sep=0.6cm]
\lstick{$C$}            & \gate{\alpha\ket{0}+\beta\ket{1}}       & \ctrl{3}                      & \ctrl{2}                              & \ctrl{1}                                        & \qw                                 & \qw \\
\lstick{$M^{(1)}$}  & \qw                                                     & \qw                          & \qw                                  & \targ{}                                        & \ctrl{2}                             & \qw \\
\lstick{$S$}            &\qw                                                      & \qw                          & \gate[wires=1]{U_0^{(1)}/U_1^{(1)}}   &   \gate[wires=1]{F_0^{(1)}/F_1^{(1)}}    & \qw                                 & \qw \\
\lstick{$P$}            & \gate{\gamma\ket{0}+\delta \ket{1}}       & \targ{}                   & \qw                                    &   \ctrl{-1}                                & \gate[wires=1]{V_0^{(1)}/V_1^{(1)}}  & \qw
\end{quantikz}
}
\caption{Full quantum circuits of the first QDM iteration 
(with either uncorrelated (top) or correlated (bottom) 
initialization step included). 
Feedback $ \mathcal{F}^{(1)}$ acts on $S$ using the initial 
policy $P$ before the policy update $\mathcal{V}^{(1)}$ is applied.}
\label{fig:1st-iter}
\end{figure}

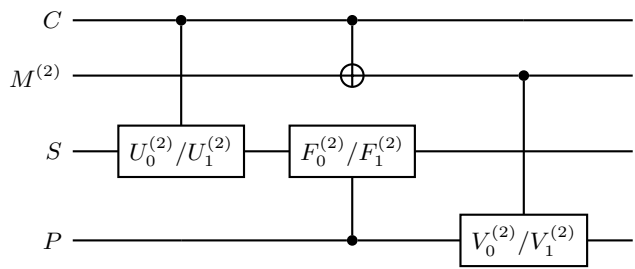
\begin{figure}[H]
\centering
\scalebox{1}{\begin{quantikz}[row sep=0.5cm, column sep=0.6cm]
\lstick{$C$}    & \ctrl{2}                                                            & \ctrl{1}                                        & \qw                                 & \qw \\
\lstick{$M^{(2)}$}   & \qw                                                        & \targ{}                                        & \ctrl{2}                             & \qw \\
\lstick{$S$}     & \gate[wires=1]{U_0^{(2)}/U_1^{(2)}}              &  \gate[wires=1]{F_0^{(2)}/F_1^{(2)}}    & \qw                                 & \qw \\
\lstick{$P$}    & \qw                                                                 &   \ctrl{-1}                                & \gate[wires=1]{V_0^{(2)}/V_1^{(2)}}  & \qw
\end{quantikz}
}
\caption{Full quantum circuit of the second QDM iteration. Feedback 
$ \mathcal{F}^{(2)}$ acts on $S$ using the policy state after the first iteration.}
\label{fig:2nd-iter}
\end{figure}

Iterating for $n$ steps produces the total evolution,
\begin{equation}
  \label{eq:QDM_total_evolution}
  \ket{\Psi_{\rm global}}
  \equiv 
  \ket{\Psi}^{(n)}
  \equiv
  \mathcal{W} \ket{\Psi}^{(0)}
  = 
  \mathcal{W}^{(n)} \mathcal{W}^{(n-1)} \cdots  \mathcal{W}^{(1)} \, \ket{\Psi}^{(0)},
\end{equation}
where $\ket{\Psi}^{(0)}$ denotes either the correlated or 
uncorrelated initialization depending on context, and 
the operator order reflects the usual quantum-circuit 
convention, with operations written on the right acting 
first. Figs.\ \ref{fig:1st-iter} and \ref{fig:2nd-iter} show 
the full circuits for the first and second iterations.

This construction illustrates a fundamental principle of the QDM where 
each component plays a distinct yet interdependent role in maintaining 
a coherent, adaptive evolution. 
At each step, the control qubit $C$ seeds the branch structure into 
the corresponding memory register $M^{(k)}$, 
while the system $S$ evolves under the branch-dependent unitary $U^{(k)}_b$. 
Immediately afterward, $S$ receives coherent feedback from the current 
state of the policy register $P$ via $\mathcal{F}^{(k)}$, 
so that the system evolution is dynamically informed by past policy adaptations. 
Finally, the policy register itself is updated via $\mathcal{V}^{(k)}$, 
encoding the outcome of the iteration and shaping future feedback. 

An important feature of this update cycle is the presence of what we 
term a \emph{reflective channel}. Because the policy register $P$ is updated 
both by memory-derived signals and by its own prior configuration, part 
of the machine's strategy space is continuously reapplied to itself. 
This provides an internal mechanism for \emph{adaptation} of the 
policy within the unitary dynamics. We emphasize, however, that 
under the restricted design adopted in the main text (fresh memories 
initialized in $\ket{0}$ and no coherent modification of $C$), this 
reflective adaptation acts only on $P$ (and thereby on subsequent 
feedback to $S$) and does not directly alter the control register $C$ 
or the orthogonal memory labels that define branches 
(Section \ref{sec:branches}). In other words, reflective channels 
produce internal, self-referential adaptation of policy and 
feedback, but they are not being used here to ``reinforce'' 
branch amplitudes by changing $C$ itself (however, cf.\ 
Section \ref{sec:reinforcement} for a quick discussion of extended
model).\footnote{The term ``reflective 
channel'' is introduced here as a structural and operational construct 
within the QDM framework. It denotes a process by which components 
of the QDM feed back into themselves under the iterative update map 
$\mathcal{V}^{(k)}$, encoding self-referential modes of action. 
Distinct from standard environment-based channels in quantum 
information theory, reflective channels provide a minimal form of 
internal adaptation within coherent, history-dependent quantum 
processes. The system monitors and reapplies selected aspects of 
its own evolution over discrete time. The term is purely algebraic, 
formalizing self-referential operations within coherent, 
history-dependent quantum processes. Compare to the term
\emph{reflection} used in \cite{Dunjko2015}.}

Through repeated iterations, this process establishes a web of correlations. 
$C$ maintains a superposition of alternatives, each alternative leaves a record in 
a fresh register $M^{(k)}$, 
$S$ evolves under both branch-conditioned operations and policy-informed 
feedback, 
and $P$ continually adapts to the accumulated history. 
Even without explicit measurement, the architecture ensures that 
the entire evolution preserves coherence, 
allowing the QDM to carry information about both the instantaneous 
branch structure and the cumulative history of policy interactions. 
One can thus view the machine as performing a coherent, 
memory-guided, adaptive computation in which past outcomes influence 
future dynamics in a fully unitary fashion.

To make the iterative structure fully explicit, Figs.~\ref{fig:qdm-3iter-uncorr} 
and \ref{fig:qdm-3iter} 
show the case of three complete iterations (with uncorrelated and 
correlated initialization, respectively). 
Each round includes: (i) a branch-dependent unitary $U_b^{(k)}$ 
applied to the system under control of $C$, 
(ii) a coherent memory deposit from $C$ into a fresh memory 
register $M^{(k)}$, 
and (iii) a conditional update $V^{(k)}_b$ of the policy register $P$. 
In addition, in every round, the policy register itself acts back on 
the system, providing feedback via controlled operations $F_p^{(k)}$, 
with the first iteration already including $\mathcal{F}^{(1)}$ from 
the initial policy state. 
These diagrams therefore illustrate the entanglement structure of 
the machine after three iterations in explicit form.
\begin{figure}[H]
\centering
\scalebox{0.7}{\begin{quantikz}[row sep=0.4cm, column sep=0.6cm]
\lstick{$C$} 
  & \gate{\alpha\ket{0}+\beta\ket{1}} & \ctrl{4} & \ctrl{1} & \qw 
  & \ctrl{4} & \ctrl{2} & \qw 
  & \ctrl{4} & \ctrl{3} & \qw & \qw \\
\lstick{$M^{(1)}$} 
  & \qw & \qw & \targ{} & \ctrl{4} 
  & \qw & \qw & \qw 
  & \qw & \qw & \qw & \qw \\
\lstick{$M^{(2)}$} 
  & \qw & \qw & \qw & \qw 
  & \qw & \targ{} & \ctrl{3}  
  & \qw & \qw & \qw & \qw \\
\lstick{$M^{(3)}$} 
  & \qw  & \qw & \qw & \qw 
  & \qw & \qw & \qw 
  & \qw & \targ{} & \ctrl{2}  & \qw\\
\lstick{$S$} 
  & \qw  & \gate[wires=1]{U^{(1)}_0 / U^{(1)}_1} & \gate[wires=1]{F^{(1)}_0 / F^{(1)}_1}  & \qw 
  & \gate[wires=1]{U^{(2)}_0 / U^{(2)}_1} &  \gate[wires=1]{F^{(2)}_0 / F^{(2)}_1}      & \qw 
  & \gate[wires=1]{U^{(3)}_0 / U^{(3)}_1} & \gate[wires=1]{F^{(3)}_0 / F^{(3)}_1}    & \qw  & \qw\\
\lstick{$P$} 
  &\gate{\gamma\ket{0}+\delta\ket{1}}   & \qw & \ctrl{-1} & \gate[wires=1]{V^{(1)}_0 / V^{(1)}_1} 
  & \qw & \ctrl{-1} & \gate[wires=1]{V^{(2)}_0 / V^{(2)}_1} 
  & \qw& \ctrl{-1}  & \gate[wires=1]{V^{(3)}_0 / V^{(3)}_1} & \qw
\end{quantikz}
}
\caption{Explicit three-iteration Quantum Deliberating Machine (QDM) 
circuit (shown here for uncorrelated initialization). 
Each iteration applies a branch-dependent system 
unitary $U^{(k)}_b$, copies the control 
into a fresh memory register $M^{(k)}$, and updates the policy register 
$P$ via $V^{(k)}_b$. From the first round onward, $P$ acts back on 
the system $S$ via controlled feedback operations $F^{(k)}_p$ 
(shown as controls from $P$ to $S$), with feedback using 
the current policy state before it is updated. In the restricted design 
used in the main text (fresh $M^{(k)}$ initialized in $\ket{0}$ and $C$ not 
coherently modified between writes) the branching multiplicity per 
iteration remains at most two, although the memory-string labels grow 
with each iteration.}
\label{fig:qdm-3iter-uncorr}
\end{figure}
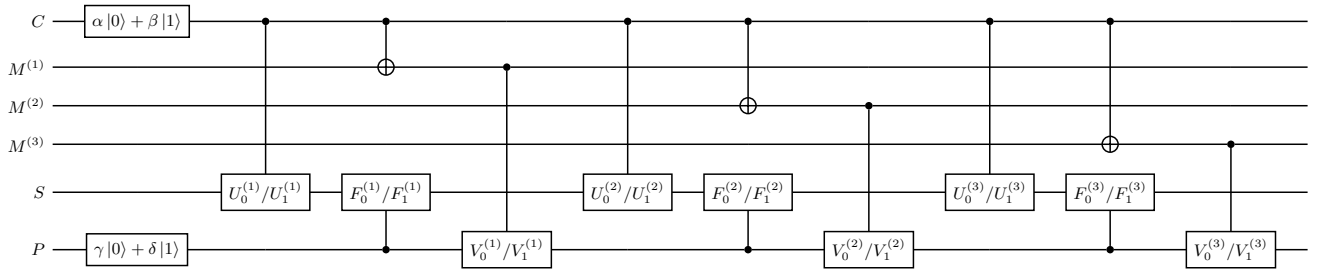

\begin{figure}[H]
\centering
\scalebox{0.7}{\begin{quantikz}[row sep=0.4cm, column sep=0.6cm]
\lstick{$C$} 
  & \gate{\alpha\ket{0}+\beta\ket{1}} & \ctrl{5}& \ctrl{4} & \ctrl{1} & \qw 
  & \ctrl{4} & \ctrl{2} & \qw 
  & \ctrl{4} & \ctrl{3} & \qw & \qw \\
\lstick{$M^{(1)}$} 
  & \qw & \qw & \qw & \targ{} & \ctrl{4} 
  & \qw & \qw & \qw 
  & \qw & \qw & \qw & \qw \\
\lstick{$M^{(2)}$} 
  & \qw & \qw & \qw & \qw & \qw 
  & \qw & \targ{} & \ctrl{3}  
  & \qw & \qw & \qw & \qw \\
\lstick{$M^{(3)}$} 
  & \qw & \qw & \qw & \qw & \qw 
  & \qw & \qw & \qw 
  & \qw & \targ{} & \ctrl{2}  & \qw\\
\lstick{$S$} 
  & \qw & \qw & \gate[wires=1]{U^{(1)}_0 / U^{(1)}_1} & \gate[wires=1]{F^{(1)}_0 / F^{(1)}_1}  & \qw 
  & \gate[wires=1]{U^{(2)}_0 / U^{(2)}_1} &  \gate[wires=1]{F^{(2)}_0 / F^{(2)}_1}      & \qw 
  & \gate[wires=1]{U^{(3)}_0 / U^{(3)}_1} & \gate[wires=1]{F^{(3)}_0 / F^{(3)}_1}    & \qw  & \qw\\
\lstick{$P$} 
  &\gate{\gamma\ket{0}+\delta\ket{1}}  & \targ{} & \qw & \ctrl{-1} & \gate[wires=1]{V^{(1)}_0 / V^{(1)}_1} 
  & \qw & \ctrl{-1} & \gate[wires=1]{V^{(2)}_0 / V^{(2)}_1} 
  & \qw& \ctrl{-1}  & \gate[wires=1]{V^{(3)}_0 / V^{(3)}_1} & \qw
\end{quantikz}
}
\caption{Explicit three-iteration QDM circuit, here shown for correlated initialization; 
cf.\ Fig.\ \ref{fig:qdm-3iter-uncorr}.}
\label{fig:qdm-3iter}
\end{figure}

\section{Branches and Orthogonality of Memory Records}
\label{sec:branches}

A subtle but important question concerns what precisely constitutes a 
\emph{branch} in the QDM. It is reasonable to distinguish between 
branches per se (\emph{main branches}) and their respective 
sub-branches. To understand the idea, let us consider the global 
state of the machine after the first iteration, assuming for definiteness 
\emph{uncorrelated} initialization as an example. Acting on 
$\ket{\Psi}_{\rm uncorr}^{(0)}$, Eq.\ (\ref{eq:uncorrelated-initialization}), 
with 
$\mathcal{W}^{(1)} = \mathcal{V}^{(1)} \,  
\mathcal{F}^{(1)} \, \mathrm{CNOT}_{C\to M^{(1)}} \, 
\mathcal{U}^{(1)}$, we find,
\begin{align}
\ket{\Psi}_{\rm uncorr}^{(1)}
&=
\mathcal{W}^{(1)}\ket{\Psi}_{\rm uncorr}^{(0)}
\nonumber \\
&=
\alpha\ket{0}_{C}\ket{0}_{M^{(1)}}
\underbrace{
\left[
\gamma F_0^{(1)}U_0^{(1)}\ket{\psi}_S^{(0)} \otimes V_0^{(1)}\ket{0}_P 
+
\delta  F_1^{(1)}U_0^{(1)}\ket{\psi}_S^{(0)} \otimes V_0^{(1)}\ket{1}_P 
\right]
}_{{\rm (main)}\; \alpha{\rm-branch}}
\nonumber \\
&\quad+
\beta\ket{1}_{C}\ket{1}_{M^{(1)}}
\underbrace{
\left[
\gamma F_0^{(1)}U_1^{(1)}\ket{\psi}_S^{(0)}\otimes V_1^{(1)}\ket{0}_P 
+
\delta  F_1^{(1)}U_1^{(1)}\ket{\psi}_S^{(0)} \otimes V_1^{(1)}\ket{1}_P 
\right]
}_{{\rm (main)}\; \beta{\rm-branch}}
\nonumber \\
&=
\alpha\ket{0}_{C}\ket{0}_{M^{(1)}}
\left\{
\underbrace{
\Big[
\Big(
\gamma V_{0,00}^{(1)} F_0^{(1)}
+
\delta V_{0,01}^{(1)} F_1^{(1)}
\Big) \,
U_0^{(1)}\ket{\psi}_S^{(0)}
\Big]  \otimes  \ket{0}_P 
}_{{\rm primary}\; \alpha{\rm-subbranch}}
\right.
\nonumber \\
& \left. \qquad \qquad \qquad \qquad
+
\underbrace{
\Big[
\Big(
\gamma V_{0,10}^{(1)} F_0^{(1)}
+
\delta V_{0,11}^{(1)} F_1^{(1)}
\Big) \,
U_0^{(1)}\ket{\psi}_S^{(0)}
\Big] \otimes  \ket{1}_P 
}_{{\rm secondary}\; \alpha{\rm-subbranch}}
\right\}
\nonumber \\
&\quad +
\beta\ket{1}_{C}\ket{1}_{M^{(1)}}
\left\{
\underbrace{
\Big[
\Big(
\gamma V_{1,00}^{(1)} F_0^{(1)}
+
\delta V_{1,01}^{(1)} F_1^{(1)}
\Big) \,
U_1^{(1)}\ket{\psi}_S^{(0)}
\Big]  \otimes \ket{0}_P 
}_{{\rm secondary}\; \beta{\rm-subbranch}}
\right.
\nonumber \\
& \left. \quad \qquad \qquad \qquad \qquad 
+
\underbrace{
\Big[
\Big(
\gamma V_{1,10}^{(1)} F_0^{(1)}
+
\delta V_{1,11}^{(1)} F_1^{(1)}
\Big) \,
U_1^{(1)}\ket{\psi}_S^{(0)}
\Big] \otimes \ket{1}_P 
}_{{\rm primary}\; \beta{\rm-subbranch}}
\right\},
\end{align}
where $V_{j,\ell q}^{(1)}\equiv \langle \ell | V_j^{(1)} | q\rangle$ 
denotes the $(\ell q)$ matrix element of the operator 
$V_j^{(1)}$, $j=0,1$, in the computational basis, acting on the 
policy register (so that, e.\ g., 
$V_j^{(1)}\ket{0}_P=V_{j,00}^{(1)}\ket{0}_P + V_{j,10}^{(1)}\ket{1}_P $,
etc.). 

Similarly, after the second iteration with 
$\mathcal{W}^{(2)} = \mathcal{V}^{(2)} \,  
\mathcal{F}^{(2)} \, \mathrm{CNOT}_{C\to M^{(2)}} \, 
\mathcal{U}^{(2)}$, we get,
\begin{align}
\ket{\Psi}_{\rm uncorr}^{(2)}
&=
\mathcal{W}^{(2)}\mathcal{W}^{(1)}\ket{\Psi}_{\rm uncorr}^{(0)}
\nonumber \\
\nonumber \\
&=
\alpha\ket{0}_{C}\ket{0}_{M^{(1)}}\ket{0}_{M^{(2)}}
\left\{
\Big[
F_0^{(2)}U_0^{(2)}\Big(
\gamma V_{0,00}^{(1)} F_0^{(1)}
+
\delta V_{0,01}^{(1)} F_1^{(1)}
\Big) \,
U_0^{(1)}\ket{\psi}_S^{(0)}
\Big]  \otimes  V_0^{(2)}\ket{0}_P 
\right.
\nonumber \\
& \left. \qquad \qquad \qquad \qquad \qquad \qquad 
+
\Big[
F_1^{(2)}U_0^{(2)}\Big(
\gamma V_{0,10}^{(1)} F_0^{(1)}
+
\delta V_{0,11}^{(1)} F_1^{(1)}
\Big) \,
U_0^{(1)}\ket{\psi}_S^{(0)}
\Big] \otimes  V_1^{(2)}\ket{1}_P 
\right\}
\nonumber \\
&\quad +
\beta\ket{1}_{C}\ket{1}_{M^{(1)}}\ket{1}_{M^{(2)}}
\left\{
\Big[
F_0^{(2)}U_1^{(2)}\Big(
\gamma V_{1,00}^{(1)} F_0^{(1)}
+
\delta V_{1,01}^{(1)} F_1^{(1)}
\Big) \,
U_1^{(1)}\ket{\psi}_S^{(0)}
\Big]  \otimes V_0^{(2)}\ket{0}_P 
\right.
\nonumber \\
& \left.  \quad \qquad \qquad \qquad \qquad \qquad \qquad 
+
\Big[
F_1^{(2)}U_1^{(2)}\Big(
\gamma V_{1,10}^{(1)} F_0^{(1)}
+
\delta V_{1,11}^{(1)} F_1^{(1)}
\Big) \,
U_1^{(1)}\ket{\psi}_S^{(0)}
\Big] \otimes V_1^{(2)}\ket{1}_P 
\right\},
\end{align}
and so on.

Thus, operationally, (main) branches are defined by the 
memory registers, while the corresponding subbranches are 
distinguished by the computational basis states of the policy 
register $P$. Each iteration of the machine copies the value of the 
control register $C$ into a fresh memory register $M^{(k)}$, 
initialized in $\ket{0}_{M^{(k)}}$. After $n$ iterations, the concatenation 
of all memory slots yields a bit string, 
$\mathbf b = (b^{(1)} b^{(2)} \ldots b^{(n)})$. Distinct strings 
correspond to orthogonal basis states of the global memory space. 
Projecting the global state onto such a basis component,
\[
\ket{\Psi_{\mathbf b}} = 
\frac{\big( \Pi_{\mathbf b} \otimes I_{\text{rest}} \big)\ket{\Psi_{\rm global}}}
{
\sqrt{
\bra{\Psi_{\rm global}} 
\big( \Pi_{\mathbf b} \otimes I_{\text{rest}} \big)\ket{\Psi_{\rm global}} 
}
},
\]
yields the pure normalized substate associated with branch $\mathbf b$, where 
$\Pi_{\mathbf b} = \ket{\mathbf b}\bra{\mathbf b}$ acts on the memory 
registers only.

\vskip5pt

{\it Definition (main branch).} A main branch is the component of 
the global state associated with a fixed computational-basis string 
of all memory registers (projector $\Pi_{\mathbf b}$). 
Branches are orthogonal because different memory strings 
are orthogonal. Registers not recorded into memory (e.g. policy) 
do not contribute to branch multiplicity, only to intra-branch 
superpositions (subbranches).

\vskip5pt

We define main branches by memory strings because memory 
registers are designed to be fresh, computational-basis records. 
If other registers (e.\ g., $P$ or even $S$) became orthogonal 
across histories (become correlated with histories in an orthogonal 
way), they could in principle generate equivalent branch 
decompositions, and therefore could also be used as branch labels.
From this perspective, two conclusions immediately follow:

\begin{enumerate}
    \item Branches are labeled by \emph{orthogonal memory records}. 
    Once the memories differ in computational-basis content, the 
    corresponding substates are orthogonal, even if the system $S$ or  
    policy $P$ remain in coherent superposition.
    \item A branch does \emph{not} require the system $S$ to be in a 
    basis state. Within each branch, $S$ (and possibly $P$) can occupy 
    an arbitrary pure state, including a superposition or entangled 
    configuration. What makes two branches distinct is solely the 
    orthogonality of their memory records.
\end{enumerate}

This definition also clarifies why in the minimal one-qubit control 
architecture only two branches ever appear under our basic restricted 
design (see explicit examples below). If $C$ is a single qubit superposed 
as $ \alpha\ket{0}_C+\beta\ket{1}_C$ and $C$ is not coherently 
modified between writes (each $M^{(k)}$ initialized in $\ket{0}$ 
and written by CNOT from $C$), then every memory slot 
records the same value and the memory space contains only 
the two strings $00\ldots0$ and $11\ldots1$, yielding 
a GHZ-like structure with only two orthogonal memory 
strings. To generate more than two branches, one must 
either enlarge the control register (so that each iteration 
can record one of several orthogonal labels),\footnote{For 
example, if the control register $C$ is taken to be two 
qubits in superposition, and both are copied into 
each memory slot, then one obtains four orthogonal memory 
strings, corresponding to the four basis states of $C$. The 
structure is otherwise identical to the one-qubit case, but the 
branching space is larger.} or allow the control 
value itself to evolve coherently between iterations (so that 
different histories deposit different memory strings, as in our
extended model discussed in Sec.\ \ref{sec:reinforcement}). Both are 
natural generalizations that preserve the branch-based semantics 
of the QDM.

\section{Example 1: Pauli-based flips}

To illustrate the QDM dynamics in detail, and to make the algebra 
explicit and easy to follow, we choose a small set of simple unitaries 
that are standard in pedagogical examples. These are intentionally 
discrete, so that applying them to computational-basis states yields 
basis states (no extraneous trigonometric algebra), which keeps 
the amplitude flow transparent. Let the controlled unitaries be
\begin{align}
\mathcal{U}^{(k)} &= \begin{cases}
I, & \text{if } C=0, \\
X, & \text{if } C=1,
\end{cases}
\end{align}
so that the control register $C$ applies either the identity or 
the Pauli $X$ to the system $S$.  
For the policy-dependent feedback, we take
\begin{align}
\mathcal{F}^{(k)} &= \begin{cases}
I, & \text{if } P=0, \\
Z, & \text{if } P=1,
\end{cases}
\end{align}
and for the memory-controlled policy updates,
\begin{align}
\mathcal{V}^{(k)} &= \begin{cases}
I, & \text{if } M^{(k)}=0, \\
X, & \text{if } M^{(k)}=1.
\end{cases}
\end{align}
Thus, branch 0 leaves the system alone, branch 1 flips the system 
with Pauli $X$; when the memory indicates branch 1 we flip 
the policy with Pauli $X$; and the policy, if in state $\ket{1}_P$, 
applies a Pauli $Z$ to the system as feedback. These choices are 
simple yet nontrivial. They allow the policy to change in response 
to the recorded branch and allow that changed policy to influence 
later system dynamics via $Z$.

\begin{figure}[H]
\centering
\scalebox{0.95}{\begin{quantikz}[row sep=0.4cm, column sep=0.6cm]
\lstick{$C$} 
& \gate{H}   & \ctrl{4} & \ctrl{1} & \qw 
  & \ctrl{4} & \ctrl{2} & \qw 
  & \ctrl{4} & \ctrl{3} & \qw & \qw \\
\lstick{$M^{(1)}$} 
& \qw    & \qw & \targ{} & \ctrl{4} & \qw & \qw 
  & \qw & \qw & \qw & \qw \\
\lstick{$M^{(2)}$} 
& \qw   & \qw & \qw & \qw 
  & \qw & \targ{} & \ctrl{3}  
  & \qw & \qw & \qw & \qw \\
\lstick{$M^{(3)}$} 
& \qw   & \qw & \qw & \qw 
  & \qw & \qw & \qw 
  & \qw & \targ{} & \ctrl{2}  & \qw\\
\lstick{$S$} 
& \qw   & \gate[wires=1]{I/X} & \gate[wires=1]{I/Z}  & \qw 
  & \gate[wires=1]{I/X} &  \gate[wires=1]{I/Z}      & \qw 
  & \gate[wires=1]{I/X} & \gate[wires=1]{I/Z}    & \qw  & \qw\\
\lstick{$P$} 
& \qw   & \qw & \ctrl{-1} & \gate[wires=1]{I/X} 
  & \qw & \ctrl{-1} & \gate[wires=1]{I/X} 
  & \qw& \ctrl{-1}  & \gate[wires=1]{I/X} & \qw
\end{quantikz}
}
\caption{Explicit QDM circuit for the simple Pauli-based three-step scenario.}
\label{fig:qdm-3iter-SIMPLE}
\end{figure}
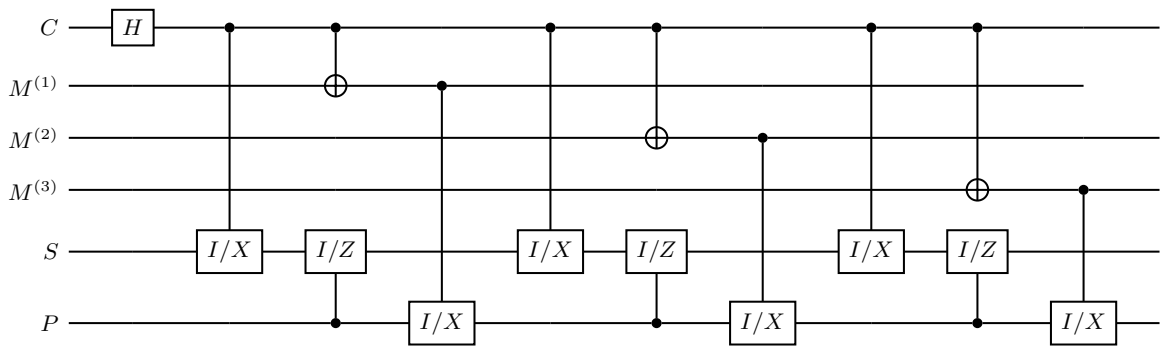

We initialize the registers as
\begin{equation}
\ket{\Psi_0} = \frac{1}{\sqrt{2}}(\ket{0}_C 
+ \ket{1}_C)\ket{000}_M\ket{0}_S\ket{0}_P,
\end{equation}
that is, the control starts in an equal superposition, the system and 
memory slots are in computational zero, and the policy begins in 
$\ket{0}_P$ (uncorrelated initialization). 
We now apply the three iterations step by step and 
write the full global state after each iteration. The register ordering 
we use in the kets below is
\[
  \big( C,\, M^{(1)},\, M^{(2)},\, M^{(3)},\, S,\, P \big),
\]
which matches the wire ordering (top to bottom) in the explicit circuit 
diagram depicted in Fig.~\ref{fig:qdm-3iter-uncorr}. 
The full three-step diagram corresponding to our scenario is depicted 
in Fig.\ \ref{fig:qdm-3iter-SIMPLE}.

\paragraph{First iteration.}
Apply $\mathcal{U}^{(1)}$: if $C=0$, $S$ unchanged; if $C=1$, $S$ flipped. 
Then deposit $C$ into $M^{(1)}$, and apply $V^{(1)}$. The state becomes
\begin{equation}
\ket{\Psi_1}=\frac{1}{\sqrt{2}}\Big(
   \ket{0}_C\ket{0}_{M^{(1)}}\ket{0}_{M^{(2)}}\ket{0}_{M^{(3)}}\ket{0}_S\ket{0}_P
 + \ket{1}_C\ket{1}_{M^{(1)}}\ket{0}_{M^{(2)}}\ket{0}_{M^{(3)}}\ket{1}_S\ket{1}_P
\Big).
\end{equation}
Note that in the first iteration feedback is trivial because $P$ was initialized in $\ket{0}_P$.

\paragraph{Second iteration.}
Apply $\mathcal{U}^{(2)}$,
\begin{align*}
&\ket{0}_C\ket{0}_{M^{(1)}}\cdots\ket{0}_S 
\mapsto \ket{0}_C\ket{0}_{M^{(1)}}\cdots\ket{0}_S, \\
&\ket{1}_C\ket{1}_{M^{(1)}}\cdots\ket{1}_S 
\mapsto \ket{1}_C\ket{1}_{M^{(1)}}\cdots X\ket{1}_S.
\end{align*}
Thus,
\[
\ket{\Psi} = \frac{1}{\sqrt{2}}\Big( 
   \ket{0}_C\ket{0}_{M^{(1)}}\ket{0}_{M^{(2)}}\ket{0}_{M^{(3)}}\ket{0}_S\ket{0}_P
 + \ket{1}_C\ket{1}_{M^{(1)}}\ket{0}_{M^{(2)}}\ket{0}_{M^{(3)}}\ket{0}_S\ket{1}_P
\Big).
\]
Now deposit $C$ into $M^{(2)}$,
\[
\ket{\Psi} = \frac{1}{\sqrt{2}}\Big( 
   \ket{0}_C\ket{0}_{M^{(1)}}\ket{0}_{M^{(2)}}\ket{0}_{M^{(3)}}\ket{0}_S\ket{0}_P
 + \ket{1}_C\ket{1}_{M^{(1)}}\ket{1}_{M^{(2)}}\ket{0}_{M^{(3)}}\ket{0}_S\ket{1}_P
\Big).
\]
Apply feedback $F^{(2)}$: in both branches $S=\ket{0}$, so no change.  
Now apply $V^{(2)}$: if $M^{(2)}=0$, $P$ unchanged; if $M^{(2)}=1$, 
apply $X$ to $P$, flipping $\ket{1}_P$ to $\ket{0}_P$.  
Thus,
\begin{equation}
\ket{\Psi_2}=\frac{1}{\sqrt{2}}\Big(
   \ket{0}_C\ket{0}_{M^{(1)}}\ket{0}_{M^{(2)}}\ket{0}_{M^{(3)}}\ket{0}_S\ket{0}_P
 + \ket{1}_C\ket{1}_{M^{(1)}}\ket{1}_{M^{(2)}}\ket{0}_{M^{(3)}}\ket{0}_S\ket{0}_P
\Big).
\end{equation}

\paragraph{Third iteration.}
Apply $\mathcal{U}^{(3)}$: in the $C=0$ branch, $S=\ket{0}$ unchanged; 
in the $C=1$ branch, $S=\ket{0}$ flips to $\ket{1}$.  
Deposit $C$ into $M^{(3)}$,
\[
\ket{\Psi}=\frac{1}{\sqrt{2}}\Big(
   \ket{0}_C\ket{0}_{M^{(1)}}\ket{0}_{M^{(2)}}\ket{0}_{M^{(3)}}\ket{0}_S\ket{0}_P
 + \ket{1}_C\ket{1}_{M^{(1)}}\ket{1}_{M^{(2)}}\ket{1}_{M^{(3)}}\ket{1}_S\ket{0}_P
\Big).
\]
Apply $F^{(3)}$: since $P=0$ in both branches, no effect on $S$.  
Now apply $V^{(3)}$: if $M^{(3)}=0$, $P=\ket{0}$ unchanged; 
if $M^{(3)}=1$, apply $X$, flipping $P$ to $\ket{1}$.  
Thus the final state after three iterations is
\begin{equation}
\ket{\Psi_3}=\frac{1}{\sqrt{2}}\Big(
   \ket{0}_C\ket{0}_{M^{(1)}}\ket{0}_{M^{(2)}}\ket{0}_{M^{(3)}}\ket{0}_S\ket{0}_P
 + \ket{1}_C\ket{1}_{M^{(1)}}\ket{1}_{M^{(2)}}\ket{1}_{M^{(3)}}\ket{1}_S\ket{1}_P
\Big).
\label{eq:Psi3}
\end{equation}
This is a generalized GHZ-like state across the six registers (control, three 
memories, system, policy). It is a compact, exact expression and it matches 
our three-step circuit diagram. Each memory slot records the corresponding 
control bit, and the system and policy have been adaptively modified so that, 
in this simple example, they end up carrying the same classical bit as 
the original control, but in a fully coherent and entangled way.

Equation~(\ref{eq:Psi3}) is instructive. The reduced state of any single 
memory qubit (say $M^{(1)}$) is
\begin{equation}
  \rho_{M^{(1)}} = \mathrm{Tr}_{\text{others}} \big( \ket{\Psi_3}\bra{\Psi_3} \big)
  = \frac{1}{2}\ket{0}\bra{0} + \frac{1}{2}\ket{1}\bra{1},
\end{equation}
a classical mixture. Although the global state is pure and fully 
coherent, the state of any particular memory register, when viewed 
alone, appears classical and diagonal in the computational basis. 
The no-cloning theorem is not violated here, because we 
did not produce independent pure copies of the unknown 
control amplitudes, but rather an entangled GHZ state that contains classical 
correlations in its reduced marginals (refer to Sec.\ \ref{sec:marginals} 
for fuller discussion). If we decided to measure, for example, the memory 
registers, the measurement outcome would collapse the GHZ state and 
project the other registers accordingly. That is, measuring all three 
memories and obtaining the string $b b b$ would collapse the global state 
onto the corresponding branch substate with system and policy labels equal 
to $\ket{b}$.

Finally, this example shows how simple Pauli gates and CNOTs 
suffice to produce a rich adaptive, coherent dynamics, 
in which policy changes in response to memory and then 
(potentially) feeds back into system evolution. In our specific 
scenario the feedback had trivial action in two rounds because 
the system happened to be in an eigenstate of the feedback 
operator; with other choices 
(say, different initial states or different feedback unitaries) 
the feedback would produce nontrivial entangling dynamics 
between $S$ and $P$. This example therefore 
serves both as a demonstration and as a template for more 
complicated situations. One may substitute any desired 
$U_b^{(k)}$, $V_b^{(k)}$, and $F_b^{(k)}$, and carry out
the same operator-composition steps to obtain non-trivial 
global state after $n$ iterations.

\section{Example 2: Rotations with and without Feedback}

In this section we present a more complicated, three-iteration 
example in two variants:
\begin{itemize}
  \item[] {\bf Version A (No feedback):} the policy $P$ is present but no feedback 
  to the system $S$ is applied (all $F^{(k)}=I$). This serves as a baseline.
  \item[] {\bf Version B (With feedback):} the policy $P$ is correlated with 
  the control $C$, and each iteration applies a small policy-controlled extra 
  rotation (an ``extra push'') on the system $S$.
\end{itemize}

Our operational scenario is chosen as follows. The system qubit $S$ starts 
in $\ket{0}$ and then evolves along two parallel paths, rotating around 
the $x$-axis in opposite directions toward $\ket{1}$ by angles $\pm \theta$ 
at each iteration. The policy $P$ provides a small, branch-dependent 
``extra push'' at each step, corresponding to an additional rotation around 
the $x$-axis by $\pm \varepsilon$. 

In Version A, without feedback, the system ends up where we intuitively 
expect after three iterations, namely at $\ket{1}$. In Version B, with feedback, 
the final state depends on the branch. Our goals are to verify that the machine 
behaves as expected and to examine exactly what the feedback accomplishes.

We will choose concrete rotation angles so that the arithmetic is easy 
to follow and the physics is clear. All steps are given in the pure-state 
formalism, and the register ordering in kets is always
\[
(C,\,M^{(1)},\,M^{(2)},\,M^{(3)},\,S,\,P).
\]

\subsection{Gate choices and notation}

Let the main branch-dependent rotations around the $x$-axis be
\begin{equation}
  U^{(k)}_0 = R_x(+\theta), \qquad U^{(k)}_1 = R_x(-\theta),
\end{equation}
with $\theta = \pi/3$.
We use the standard convention
\begin{equation}
  R_x(\phi) = \exp\!\Big(-i\frac{\phi}{2} X\Big)
  = \cos\left(\frac{\phi}{2}\right) I - i \sin\left(\frac{\phi}{2}\right) X .
\end{equation}
When acting on $\ket{0}_S$ it gives,
\begin{equation}
  R_x(\phi)\ket{0} = \cos\left(\frac{\phi}{2}\right)\ket{0} 
  - i\sin\left(\frac{\phi}{2}\right)\ket{1}.
\end{equation}

We set the small extra (policy) push angle
$  \varepsilon = \pi/12$.
In Version B the feedback operator applied to $S$ is
\begin{align}
  F^{(k)} = \begin{cases}
    R_x(+\varepsilon), & \text{if } P=\ket{0},\\[4pt]
    R_x(-\varepsilon), & \text{if } P=\ket{1},
  \end{cases}
\end{align}
i.e.\ as a controlled unitary $ \mathcal{F}^{(k)}$ from $P$ onto $S$,
\begin{equation}
  \mathcal{F}^{(k)} = \ket{0}\bra{0}_P\otimes R_x(+\varepsilon)
  + \ket{1}\bra{1}_P\otimes R_x(-\varepsilon).
\end{equation}
(Version A sets all $F^{(k)}=I$.)

To ensure the policy already distinguishes branches from the first 
iteration (so feedback can act differently in iteration 1), we initialize 
$P$ correlated with $C$ by copying $C$ into $P$ before the iterations 
begin (this is the same style of coherent branch-labelling used for 
the memories).\footnote{We emphasize once again, that entangling 
$P$ with $C$ at initialization changes the \emph{content}
of the two branches (so feedback can act differently from the first
iteration), but it does not increase the number of orthogonal memory
labels when only a single control qubit is copied into every fresh
memory slot. To enlarge the branching space one must either enlarge $C$
or allow $C$ to evolve coherently between writes. Independent 
superpositions of $P$ affect intra-branch dynamics, but without 
recording $P$ into memory they do not create new branches.}
Therefore, we start with
\begin{equation}
  \ket{\Psi_{\rm init}} = \ket{+}_C \otimes \ket{0}_{M^{(1)}M^{(2)}M^{(3)}} 
  \otimes \ket{0}_S \otimes \ket{0}_P,
\end{equation}
and then perform a CNOT from $C$ to $P$ so that $P$ is correlated with $C$. 
We can equivalently think of initial joint control-policy preparation as
\begin{equation}
 \ket{\Psi'_{\rm init}} 
 =  
 \frac{1}{\sqrt{2}}\big(\ket{0}_C\ket{0}_P + \ket{1}_C\ket{1}_P\big)
  \otimes \ket{000}_M \otimes \ket{0}_S .
\end{equation}
This ensures $P$ takes the value $\ket{0}$ on the branch labelled by $C=0$ 
and $\ket{1}$ on the branch $C=1$ from the outset.

Memory updates $\mathrm{CNOT}_{C\to M^{(k)}}$ and optional 
policy updates $\mathcal{V}^{(k)}$ (if present) follow the same ordering 
as in the main text. In each iteration $k$ we apply (right-to-left),
\[
  \mathcal{W}^{(k)} = \mathcal{V}^{(k)}\;\Big(\;\mathcal{F}^{(k)}\;
  \big(\mathrm{CNOT}_{C\to M^{(k)}}\;(\mathcal{U}^{(k)}\ket{\Psi})\big)\Big),
\]
with 
\begin{equation}
\mathcal{U}^{(k)}=\ket{0}\bra{0}_C\otimes R_x(+\theta)+\ket{1}\bra{1}_C\otimes R_x(-\theta).
\end{equation}
In our example we take the policy update maps 
$\mathcal{V}^{(k)}$ to be the identity,
\begin{equation}
\mathcal{V}^{(k)}=1, \quad \forall k,
\end{equation}
so the policy remains the copied value of $C$ throughout; this isolates the effect 
of policy-controlled feedback.

\begin{figure}[H]
\centering
\scalebox{0.90}{\begin{quantikz}[row sep=0.4cm, column sep=0.6cm]
\lstick{$C$} 
& \gate{H}  & \ctrl{5} & \ctrl{4} & \ctrl{1} & \qw 
  & \ctrl{4} & \ctrl{2} & \qw 
  & \ctrl{4} & \ctrl{3} & \qw & \qw \\
\lstick{$M^{(1)}$} 
& \qw  & \qw & \qw & \targ{} & \ctrl{4} 
  & \qw & \qw & \qw 
  & \qw & \qw & \qw & \qw \\
\lstick{$M^{(2)}$} 
& \qw  & \qw & \qw & \qw & \qw 
  & \qw & \targ{} & \ctrl{3}  
  & \qw & \qw & \qw & \qw \\
\lstick{$M^{(3)}$} 
& \qw  & \qw & \qw & \qw & \qw 
  & \qw & \qw & \qw 
  & \qw & \targ{} & \ctrl{2}  & \qw\\
\lstick{$S$} 
& \qw  & \qw & \gate[wires=1]{R_x(\pm \theta)} & \gate[wires=1]{R_x(\pm \varepsilon)}  & \qw 
  & \gate[wires=1]{R_x(\pm \theta)} &  \gate[wires=1]{R_x(\pm \varepsilon)}      & \qw 
  & \gate[wires=1]{R_x(\pm \theta)} & \gate[wires=1]{R_x(\pm \varepsilon)}    & \qw  & \qw\\
\lstick{$P$} 
& \qw  &\targ{}  & \qw & \ctrl{-1} & \gate[wires=1]{{I}} 
  & \qw & \ctrl{-1} & \gate[wires=1]{{I}} 
  & \qw& \ctrl{-1}  & \gate[wires=1]{{I}} & \qw
\end{quantikz}
}
\caption{Explicit QDM circuit for the three-step rotation scenario.}
\label{fig:qdm-3iter-ROTATIONS}
\end{figure}
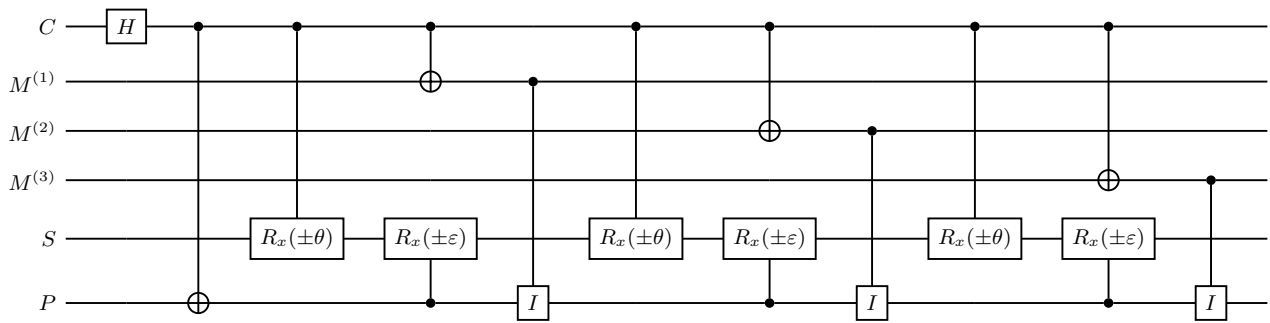

The full three-step diagram corresponding to our scenario is depicted in 
Fig.\ \ref{fig:qdm-3iter-ROTATIONS}.

\subsection{Initial global state}

After copying $C$ into $P$ we have as initial global state,
\begin{equation}
  \ket{\Psi_0}
  = \frac{1}{\sqrt{2}}\Big(
    \ket{0}_C\ket{0}_{M^{(1)}}\ket{0}_{M^{(2)}}\ket{0}_{M^{(3)}}\ket{0}_S\ket{0}_P
  + \ket{1}_C\ket{0}_{M^{(1)}}\ket{0}_{M^{(2)}}\ket{0}_{M^{(3)}}\ket{0}_S\ket{1}_P
  \Big),
\end{equation}
where we have explicitly written all $M^{(i)}$ initialized to $\ket{0}$.

\subsection{Version A: No feedback ($F^{(k)}=I$ for all $k$)}

Because feedback is absent, each iteration simply applies 
the branch-dependent rotation and then copies $C$ into the fresh 
memory slot.

\paragraph{First iteration.}
Apply $\mathcal{U}^{(1)}$ (controlled $R_x(\pm\theta)$) and 
then $\mathrm{CNOT}_{C\to M^{(1)}}$. Acting on each branch we get,
\begin{equation}
  R_x(+\theta)\ket{0}_S = \cos\left(\frac{\theta}{2}\right)\ket{0} 
  - i\sin\left(\frac{\theta}{2}\right)\ket{1},
\quad
  R_x(-\theta)\ket{0}_S = \cos\left(\frac{\theta}{2}\right)\ket{0} 
  + i\sin\frac{\theta}{2}\ket{1}.
\end{equation}
With $\theta=\pi/3$, we have $\cos(\theta/2)=\cos(\pi/6)=\sqrt{3}/2$ 
and $\sin(\theta/2)=\sin(\pi/6)=1/2$, but we will keep symbolic trig 
factors for clarity. After the controlled rotation and memory write, we have,
\begin{equation}
\ket{\Psi_1}=\frac{1}{\sqrt{2}}\Big(
  \ket{0}_C\ket{0}_{M^{(1)}}\ket{0}_{M^{(2)}}\ket{0}_{M^{(3)}}
    \;R_x(+\theta)\ket{0}_S\;\ket{0}_P
+ \ket{1}_C\ket{1}_{M^{(1)}}\ket{0}_{M^{(2)}}\ket{0}_{M^{(3)}}
    \;R_x(-\theta)\ket{0}_S\;\ket{1}_P
\Big).
\end{equation}

\paragraph{Second iteration.}
Apply $\mathcal{U}^{(2)}$ and write into $M^{(2)}$. Because rotations 
about the same axis compose additively, we get
\begin{equation}
  R_x(\pm\theta) R_x(\pm\theta) = R_x(\pm 2\theta),
\end{equation}
and the sign is the same branch sign. After second iteration and memory 
write we have,
\begin{equation}
\ket{\Psi_2}=\frac{1}{\sqrt{2}}\Big(
  \ket{0}_C\ket{0}_{M^{(1)}}\ket{0}_{M^{(2)}}\ket{0}_{M^{(3)}}
    \;R_x(+2\theta)\ket{0}_S\;\ket{0}_P
+ \ket{1}_C\ket{1}_{M^{(1)}}\ket{1}_{M^{(2)}}\ket{0}_{M^{(3)}}
    \;R_x(-2\theta)\ket{0}_S\;\ket{1}_P
\Big).
\end{equation}

\paragraph{Third iteration.}
After the third controlled rotation and memory write,
\begin{equation}
\ket{\Psi_3}=\frac{1}{\sqrt{2}}\Big(
  \ket{0}_C\ket{0}_{M^{(1)}}\ket{0}_{M^{(2)}}\ket{0}_{M^{(3)}}
    \;R_x(+3\theta)\ket{0}_S\;\ket{0}_P
+ \ket{1}_C\ket{1}_{M^{(1)}}\ket{1}_{M^{(2)}}\ket{1}_{M^{(3)}}
    \;R_x(-3\theta)\ket{0}_S\;\ket{1}_P
\Big).
\end{equation}
With $\theta=\pi/3$, we have $3\theta=\pi$, so
\begin{equation}
  R_x(\pi)\ket{0} = \cos\left(\frac{\pi}{2}\right)\ket{0} 
  - i\sin\left(\frac{\pi}{2}\right)\ket{1} = -\,i\ket{1},
\quad
  R_x(-\pi)\ket{0} = \cos\left(\frac{\pi}{2}\right)\ket{0} 
  - i\sin \left(-\frac{\pi}{2}\right)\ket{1}
  = +\,i\ket{1}.
\end{equation}
Hence both branches send the system to state $\ket{1}$ up to 
a branch-dependent phase. Explicitly,
\begin{equation}
\ket{\Psi_3}=\frac{1}{\sqrt{2}}\Big(
  -i\,\ket{0}_C\ket{000}_M\ket{1}_S\ket{0}_P
  + i\,\ket{1}_C\ket{111}_M\ket{1}_S\ket{1}_P
\Big).
\end{equation}
Finally, let's factor out the common system factor to see that $S$ has 
become disentangled,
\begin{equation}
\ket{\Psi_3} 
= 
\Big( 
\frac{1}{\sqrt{2}}(-i\ket{0}_C\ket{000}_M\ket{0}_P
  + i\ket{1}_C\ket{111}_M\ket{1}_P) 
  \Big)\;\otimes\; \ket{1}_S .
\end{equation}
Therefore in Version A (no feedback) the system qubit ends exactly 
at $\ket{1}$, as expected (probability one to find $\ket{1}$), and it becomes 
separable from the rest of the registers at the final step (the remaining 
registers keep an entangled record and phase information).

\subsection{Version B: With feedback (policy-controlled small pushes)}

Now include the policy-controlled feedback $\mathcal{F}^{(k)}$ with 
$\varepsilon=\pi/12$, and keep the policy $P$ correlated with $C$ 
(copied at the start). For simplicity we do not apply explicit further 
policy updates $\mathcal{V}^{(k)}$ (they are identity here); the value 
of $P$ is therefore constant along each branch and equals the initial 
copy of $C$.

Because rotations about the $x$-axis commute, a controlled 
$R_x(\pm\theta)$ followed by a policy-controlled 
$R_x(\pm\varepsilon)$ is equivalent (on each branch) to a single 
rotation $R_x\big(\pm(\theta+\varepsilon)\big)$.
With $P$ correlated to $C$ so that branch $C=0$ has $P=0$ 
and branch $C=1$ has $P=1$, the per-iteration net rotation on 
$S$ is:
\[
  \text{branch }0:\quad R_x\big(+(\theta+\varepsilon)\big),\qquad
  \text{branch }1:\quad R_x\big(-(\theta+\varepsilon)\big).
\]
Thus after three identical iterations the net rotation angles are additive:
\[
  \text{branch }0:\quad R_x\big(3(\theta+\varepsilon)\big),\qquad
  \text{branch }1:\quad R_x\big(-3(\theta+\varepsilon)\big).
\]

With our numerical choices $\theta=\pi/3$ and 
$\varepsilon=\pi/12$, we have,
$\theta+\varepsilon = 5\pi/12$,
$3(\theta+\varepsilon)= 5\pi/4$.
Therefore, the system states at the end of three iterations are,
\begin{align}
  \text{branch }0: \quad \ket{s_0} &= R_x\left(5\pi/4\right)\ket{0}
  = \cos\left(\frac{5\pi}{8}\right)\ket{0} - i\sin\left(\frac{5\pi}{8}\right)\ket{1},
\\  \text{branch }1: \quad \ket{s_1} &= R_x\left(-5\pi/4\right)\ket{0}
  = \cos\left(\frac{5\pi}{8}\right)\ket{0} - i\sin\left(-\frac{5\pi}{8}\right)\ket{1}
  = \cos\left(\frac{5\pi}{8}\right)\ket{0} + i\sin\left(\frac{5\pi}{8}\right)\ket{1}.
\end{align}
Thus the two branch-end states are
\begin{equation}
  \ket{s_0} \approx -0.3827\ket{0} - i(0.9239)\ket{1},
  \qquad
  \ket{s_1} \approx -0.3827\ket{0} + i(0.9239)\ket{1}.
\end{equation}

The full global state after three iterations in Version B is therefore
\begin{equation}
\ket{\Psi_3^{\rm (B)}} = \frac{1}{\sqrt{2}}\Big(
  \ket{0}_C\ket{0}_{M^{(1)}}\ket{0}_{M^{(2)}}\ket{0}_{M^{(3)}}\ket{s_0}_S\ket{0}_P
+ \ket{1}_C\ket{1}_{M^{(1)}}\ket{1}_{M^{(2)}}\ket{1}_{M^{(3)}}\ket{s_1}_S\ket{1}_P
\Big).
\end{equation}

\paragraph{Measurement statistics on $S$.}
A natural measure of the effect is the probability to find the system 
in $\ket{1}$ at the end. For either branch the probability amplitude 
of $\ket{1}$ is $\mp i\sin(5\pi/8)$ whose squared magnitude is
\[
  \Pr(1|{\text{branch}}) = |\langle 1|s_0\rangle|^2 
  = |\langle 1|s_1\rangle|^2 = \sin^2\left(\frac{5\pi}{8}\right)
  = \sin^2\left(\frac{3\pi}{8}\right)
  \approx 0.8536.
\]
Since the two branches occur with equal weight (we started 
from $\ket{+}_C$ and did not bias the amplitudes), the unconditional 
probability to measure the system in $\ket{1}$ is also
\[
  \Pr(1)_S =\frac{1}{2}|\langle 1|s_0\rangle|^2 
  + \frac{1}{2}|\langle 1|s_1\rangle|^2\approx 0.8536.
\]
Thus Version B yields about an $85.36\%$ chance to obtain 
$\ket{1}$ on $S$, whereas Version A produced $\Pr(1)_S=1$.

\paragraph{Remarks on phase and entanglement.}
Although the measurement probability for $\ket{1}$ is equal 
on the two branches, the branch states $\ket{s_0}$ and $\ket{s_1}$ 
differ by a relative phase on the $\ket{1}$ component 
(one has $-i\sin(\cdot)$, the other $+i\sin(\cdot)$). 
This means that $S$ does \emph{not} factor out as a single pure 
state in Version B the way it did in Version A. 
In Version A the system became exactly $\ket{1}$ and thus 
separable from the memories and policy; in Version B the system 
remains entangled with the branch registers (control and memories) 
because the two branch-end states are not identical up to a global 
phase. This difference is operationally relevant. In Version B coherence 
between branches is retained in cross-terms and affects any 
subsequent interference-sensitive process.

To summarize the two cases:

\begin{itemize}
  \item[] \textbf{Version A (no feedback):} Final global state factorizes 
  as (entangled record) $\otimes \ket{1}_S$. Measurement of $S$ yields 
  $\ket{1}$ with probability $1$. The two candidate processes both 
  drove $S$ by a net angle $\pm 3\theta=\pm\pi$, which propagates 
  $\ket{0}$ to $\ket{1}$ (up to phase).

  \item[] \textbf{Version B (with feedback):} With policy-controlled 
  extra pushes $\pm\varepsilon$ per iteration (and $P$ correlated 
  with $C$ from the start), the net per-branch rotation becomes 
  $\pm 3(\theta+\varepsilon)=\pm 5\pi/4$. Each branch maps 
  $\ket{0}$ to a state with $\Pr(1)\approx 0.8536$. The global state 
  remains entangled between $(C,M^{(1)},M^{(2)},M^{(3)},P)$ and $S$, 
  and $S$ is not in a pure $\ket{1}$ state. Thus feedback reduced 
  the certainty of ending at $\ket{1}$ and introduced branch-dependent 
  relative phases and entanglement.
\end{itemize}

\section{Classical Appearance of Memory Marginals}
\label{sec:marginals}

A notable feature in both examples is that the memory registers 
$M^{(k)}$ \emph{appear classical in their marginals}. This means that 
if we trace out the system $S$, policy $P$, and the other memory 
registers, the reduced density matrix of any single memory qubit will 
be diagonal in the computational basis. That is, each $M^{(k)}$ locally 
resembles a classical bit recording the branch label, even though 
the full global state remains entangled and fully coherent.\footnote{While the 
QDM’s registers collectively encode the branching history, the amount 
of classical information extractable from any single subsystem is limited 
by the Holevo bound \cite{Holevo1973}. This emphasizes that internal 
deliberation relies on entanglement and coherent correlations rather 
than classical copying of states.}

\subsection{General considerations}

Let us dive into this subject (e.g., \cite{Klyachko2006, Schilling2014, 
Lancien2017, Pironio2010, Yu2021, Hsieh2022, Wang2022}) 
and give a self-contained account that: (i) states the general facts 
about memory marginals and their ``classical appearance'', 
(ii) shows how the marginal probabilities are computed from 
the global amplitudes, (iii) applies the formulas to the three-iteration 
examples, and (iv) explains the relation of the diagonal 
memory marginals to the no-cloning theorem. We will provide 
all needed definitions and carefully spell out all important steps, 
so that the interested reader could check our calculations.

\paragraph{General definitions and notation.}  
Let the global pure state after $n$ iterations live in
\[
\ket{\Psi}^{(n)} \in \mathcal{H}_C \otimes 
\bigotimes_{k=1}^{n}\mathcal{H}_{M^{(k)}} \otimes 
\mathcal{H}_S \otimes \mathcal{H}_P .
\]
We use the computational-basis labels $c\in\{0,1\}$ for the control 
qubit at each step and $m_k\in\{0,1\}$ for the memory qubit $M^{(k)}$. 
Expand the global pure state in a basis that makes the branch labels explicit,
\begin{equation}
\label{eq:global-expansion}
\ket{\Psi}^{(n)} = \sum_{c,m_1,\dots,m_n,s,p} \alpha_{c,m_1,\dots,m_n,s,p}\; 
\ket{c}_C \otimes \ket{m_1}_{M^{(1)}}\!\otimes\cdots\otimes\ket{m_n}_{M^{(n)}} 
\otimes \ket{s}_S \otimes \ket{p}_P ,
\end{equation}
where the complex amplitudes $\alpha_{c,m_1,\dots,m_n,s,p}$ 
encode all branch, system and policy information.

The reduced (marginal) density matrix of a single memory qubit 
$M^{(k)}$ is obtained by tracing out all other subsystems,
\begin{equation}
\label{eq:rho-Mk}
\rho_{M^{(k)}} = \mathrm{Tr}_{C,\{M^{(j\neq k)}\},S,P} 
\big[\,\ket{\Psi^{(n)}}\bra{\Psi^{(n)}}\,\big].
\end{equation}

\paragraph{Why the memory marginal is diagonal (``appears classical'').}  
Because the memory qubits are written by copying the branch label 
(CNOT-like operations), the amplitudes in the expansion 
(\ref{eq:global-expansion}) are nonzero only for patterns where each 
$m_i$ matches the corresponding branch label recorded at iteration $i$. 
After tracing out the correlated degrees of freedom, any off-diagonal 
matrix elements of $\rho_{M^{(k)}}$ vanish,
\begin{equation}
\label{eq:rho-Mk-diagonal}
\rho_{M^{(k)}} = p_0^{(k)} \ket{0}\bra{0} + p_1^{(k)} \ket{1}\bra{1},
\end{equation}
with the probabilities given by summing squared magnitudes of the 
global amplitudes over all configurations that have $m_k=0$ 
(respectively, $m_k=1$):
\begin{equation}
\label{eq:pk-definition}
p_{0}^{(k)} = \sum_{\substack{c,m_1,\dots,m_n,s,p \\ m_k=0}} \big|\alpha_{c,m_1,\dots,m_n,s,p}\big|^2,
\qquad
p_{1}^{(k)} = \sum_{\substack{c,m_1,\dots,m_n,s,p \\ m_k=1}} \big|\alpha_{c,m_1,\dots,m_n,s,p}\big|^2,
\end{equation}
and $p_0^{(k)}+p_1^{(k)}=1$. This is the precise sense in which 
we say $M^{(k)}$ ``appears classical in its marginal''; the reduced 
density matrix is diagonal in the computational basis and hence 
indistinguishable from a classical probability distribution over 
$\{0,1\}$ when viewed alone.

\paragraph{When do the marginal probabilities equal the initial 
control weights?}  
If the initial control preparation is
\[
\ket{\phi}_C = \alpha\ket{0}_C + \beta\ket{1}_C,
\]
and if the protocol never performs any nontrivial amplitude-changing 
operation on the control qubit (no measurements of $C$, no unitaries 
that act nontrivially on $C$ conditioned on other amplitudes, etc.), 
then every copying operation simply imprints the control label into 
fresh memory slots without changing the weight of each branch. 
In that case,
\begin{equation}
p_0^{(k)} = |\alpha|^2,\qquad p_1^{(k)} = |\beta|^2,
\end{equation}
for all $k$. In particular, for an initial balanced superposition 
$\alpha=\beta=1/\sqrt{2}$ each memory marginal is 
$\frac{1}{2}\ket{0}\bra{0}+\frac{1}{2}\ket{1}\bra{1}$, unless some 
later operation alters the branch amplitudes.

\paragraph{Example 1 (Pauli flips, GHZ-like final state).}  
In the Pauli-based flips example the global final state after three 
iterations was (up to known phases)
\begin{equation}
\ket{\Psi}^{(3)} = \frac{1}{\sqrt{2}}\Big( \ket{0}_C\ket{000}_{M}\ket{0}_S\ket{0}_P
+ \ket{1}_C\ket{111}_{M}\ket{1}_S\ket{1}_P \Big).
\end{equation}
Tracing out $C,S,P$ and the other memory qubits yields, for each $k\in\{1,2,3\}$,
\begin{equation}
\rho_{M^{(k)}} = \frac{1}{2}\ket{0}\bra{0} + \frac{1}{2}\ket{1}\bra{1}.
\end{equation}
Here $p_0^{(k)}=p_1^{(k)}=1/2$. Each memory qubit individually looks 
exactly like a fair classical bit, while the full state is a GHZ-like entangled 
state that preserves coherence across registers.

\paragraph{Example 2 (rotations with/without feedback).}  
In the rotation examples the control is initialized in 
\begin{equation}
\ket{+}_C = \frac{1}{\sqrt{2}}(\ket{0}+\ket{1})
\end{equation}
and the protocol copies the control into memory slots. The branches 
maintain equal amplitude magnitude throughout the iterations provided 
no operation alters the amplitude of $C$ itself (and in the scenarios we 
analyzed the controlled rotations and policy-dependent rotations act 
on $S$ and $P$, not on $C$). Therefore for both Version A (no feedback) 
and Version B (policy-controlled feedback that does not act on $C$), 
the marginal probabilities are
\[
p_0^{(k)} = p_1^{(k)} = \frac{1}{2},\qquad k=1,2,3.
\]
If, instead, some operation explicitly biases or measures $C$ (or applies 
amplitude-dependent gates on $C$), then the general formulas 
(\ref{eq:pk-definition}) must be used to recompute $p_0^{(k)}$ 
and $p_1^{(k)}$.

\subsection{Memory marginals and apparent classicality in Example 2, Version B}

It is instructive to examine more closely the reduced state of the memory 
subsystem in Example 2, Version B. To see explicitly why the memory 
registers appear classical in their marginals, consider a simplified branch 
state after one iteration,
\begin{equation}
\ket{\Psi} = \frac{1}{\sqrt{2}}\Big(\ket{0}_C \ket{0}_M \ket{s_0}_S 
+ \ket{1}_C \ket{1}_M \ket{s_1}_S \Big),
\end{equation}
where $\ket{s_0}$ and $\ket{s_1}$ are arbitrary normalized superpositions 
of the system qubit basis states $\{\ket{0}_S,\ket{1}_S\}$, and $C$ and $M$ 
denote the control and memory qubits, respectively. For clarity, we ignore 
normalization factors in the derivation below; they can be easily reinstated 
if needed.

The global density matrix is
\begin{align}
\ket{\Psi}\bra{\Psi} 
&= 
\ket{0}\bra{0}_C \otimes \ket{0}\bra{0}_M \otimes \ket{s_0}\bra{s_0}_S  
+ \ket{0}\bra{1}_C \otimes \ket{0}\bra{1}_M \otimes \ket{s_0}\bra{s_1}_S 
\nonumber\\
&\quad 
+ \ket{1}\bra{0}_C \otimes \ket{1}\bra{0}_M \otimes \ket{s_1}\bra{s_0}_S 
+ \ket{1}\bra{1}_C \otimes \ket{1}\bra{1}_M \otimes \ket{s_1}\bra{s_1}_S.
\end{align}
The memory marginal is obtained by tracing out the control and system qubits,
\begin{align}
\rho_M 
&= \mathrm{Tr}_{C,S}[\ket{\Psi}\bra{\Psi}] \nonumber\\
&= \mathrm{Tr}_{C,S}\Big[ \ket{0}\bra{0}_C \otimes \ket{0}\bra{0}_M \otimes \ket{s_0}\bra{s_0}_S \Big] 
+ \mathrm{Tr}_{C,S}\Big[ \ket{0}\bra{1}_C \otimes \ket{0}\bra{1}_M \otimes \ket{s_0}\bra{s_1}_S \Big] 
\nonumber\\
&\quad 
+ \mathrm{Tr}_{C,S}\Big[ \ket{1}\bra{0}_C \otimes \ket{1}\bra{0}_M \otimes \ket{s_1}\bra{s_0}_S \Big] 
+ \mathrm{Tr}_{C,S}\Big[ \ket{1}\bra{1}_C \otimes \ket{1}\bra{1}_M \otimes \ket{s_1}\bra{s_1}_S \Big].
\end{align}
Separately evaluating each term, we get,
\begin{enumerate}
    \item $\mathrm{Tr}_{C,S}[ \ket{0}\bra{0}_C \otimes \ket{0}\bra{0}_M \otimes \ket{s_0}\bra{s_0}_S ] = \ket{0}\bra{0}_M$;
    \item $\mathrm{Tr}_{C,S}[ \ket{0}\bra{1}_C \otimes \ket{0}\bra{1}_M \otimes \ket{s_0}\bra{s_1}_S ] = (\mathrm{Tr}[\ket{0}\bra{1}_C]) (\mathrm{Tr}[\ket{s_0}\bra{s_1}_S]) \ket{0}\bra{1}_M = 0$;
    \item $\mathrm{Tr}_{C,S}[ \ket{1}\bra{0}_C \otimes \ket{1}\bra{0}_M \otimes \ket{s_1}\bra{s_0}_S ] = 0$;
    \item $\mathrm{Tr}_{C,S}[ \ket{1}\bra{1}_C \otimes \ket{1}\bra{1}_M \otimes \ket{s_1}\bra{s_1}_S ] = \ket{1}\bra{1}_M$.
\end{enumerate}
Hence the memory marginal is (restoring prefactors),
\begin{equation}
\rho_M = \frac{1}{2}\ket{0}\bra{0}_M + \frac{1}{2}\ket{1}\bra{1}_M,
\end{equation}
which is diagonal in the computational basis, independent of the specific 
superpositions $\ket{s_0}$ and $\ket{s_1}$ of the system qubit. 

The key reason the off-diagonal terms vanish is the orthogonality of 
the control qubit states, $\ket{0}_C$ and $\ket{1}_C$. Even though 
the system states $\ket{s_0}$ and $\ket{s_1}$ may have overlapping 
components, the cross terms in the density matrix always include 
$\ket{0}\bra{1}_C$ or $\ket{1}\bra{0}_C$, whose trace over $C$ is zero. 

Thus, the memory, when viewed in isolation, appears classical in its 
marginal state, carrying only classical probabilities 
$p(m_0)=p(m_1)=\frac{1}{2}$, while the full global state remains 
fully coherent and entangled. 
Coherence between the two branches resides in the correlations 
with the system qubit, thus respecting the no-cloning theorem. 
Since no arbitrary quantum state is copied, only classical branch 
labels are coherently recorded across memory qubits.

We would arrive at a similar conclusion, by tracing out the memory, and 
inspecting only the system marginal,
\begin{equation}
  \rho_S = \frac{1}{2}\ket{s_0}\bra{s_0} \;+\; \frac{1}{2}\ket{s_1}\bra{s_1}.
\end{equation}

This illustrates the general point that, even though the composite 
evolution is a perfectly coherent superposition of alternative unitary 
actions, the individual subsystems (here, system and memory) can 
each look entirely classical when reduced to their marginals. In other 
words, no local observer with access only to $M$ or only to $S$ would 
be able to detect the underlying quantum superposition. The genuinely 
quantum character is preserved only in the joint state of $(S,M)$, 
encoded in their entanglement.

Consequently, even in Example~2, Version~B, where the global state 
visibly contains both $s_0$ and $s_1$ branches, the memory qubit itself 
remains in a classical mixture with probabilities $1/2$ and $1/2$. This is 
precisely the sense in which the marginals ``appear classical'' while the 
whole system maintains quantum coherence.

\subsection{Relation to the no-cloning theorem (detailed explanation)} 
\label{sec:no-cloning-detailed}
 
The no-cloning theorem forbids a universal unitary $U$ that would perform
\begin{equation}
U\ket{\psi}\ket{0} = \ket{\psi}\ket{\psi}, \quad\text{for all }\ket{\psi}.
\end{equation}
The appearance of diagonal memory marginals does not contradict 
no-cloning because the memory qubits are not storing full copies of 
arbitrary quantum states; they store only the \emph{classical branch label} 
(the eigenvalue in the computational basis). Namely:
\begin{itemize}
\item[(a)] The global state $\ket{\Psi^{(n)}}$ remains pure and entangled. 
The memory marginals are diagonal precisely because the memory 
qubits are entangled with other subsystems in a way that destroys local 
coherences when those other subsystems are traced out.
\item[(b)] A diagonal marginal 
$\rho_{M^{(k)}} = p_0^{(k)}\ket{0}\bra{0} + p_1^{(k)}\ket{1}\bra{1}$ 
does not encode the relative phases present in $\ket{\Psi^{(n)}}$. 
Those phases and cross-branch coherences are encoded in the off-diagonal 
correlations of the full density operator $\ket{\Psi^{(n)}}\bra{\Psi^{(n)}}$, 
not in the single-qubit marginal.
\item[(c)] Because the ket of a memory qubit does not contain the full information 
about all components (amplitudes) of the ket ``belonging'' to any other register, 
it is not a clone of that register. One cannot 
reconstruct the global ket vector from any single memory qubit 
alone; reconstructing the global state requires access to joint correlations 
across multiple registers.
\end{itemize}

Thus the QDM's use of memory qubits to record branch labels is fully 
consistent with quantum mechanics. The protocol \emph{broadcasts 
classical branch information} (the eigenvalue label) into memory marginals 
while preserving global coherence and avoiding the forbidden universal 
cloning of arbitrary quantum states.

When evaluating our QDM proposal, the reader should keep the following 
operational picture in mind: each memory qubit is a local, classical-appearing 
tag for a branch, useful for conditional unitary updates and reversible feedback. 
The power of the QDM arises because these classical-appearing tags live 
inside a larger entangled state. This allows adaptive, branch-dependent 
operations that exploit global coherence (interference, phase relations, 
entanglement) while still permitting localized conditional actions based 
on the branch labels. The diagonal marginals are therefore an operational 
feature, not a contradiction of the underlying quantum character of 
the machine.

\section{Possible Extensions of Example 2}

Here are a few simple modifications of Example 2 that one 
may explore next:
\begin{enumerate}
  \item Prepare $P$ initially \emph{uncorrelated} with $C$, so that the 
  first-iteration feedback would acts similarly on both branches, and 
  only later iterations would show branch-specific feedback 
  (after $\mathcal{V}^{(k)}$ creates correlations). This variation would 
  highlight the role of initial policy entanglement.
  \item Make $\mathcal{V}^{(k)}$ nontrivial, so the policy evolves 
  depending on past recorded outcomes. This would implement a genuine 
  coherent update rule and could create more complex adaptive behavior.
  \item Replace $R_x$ rotations by rotations around different axes in 
  different iterations to generate noncommuting process-superpositions. 
  Then the order of feedback and branch rotations would matter, which 
  would result in considerably more interesting scenarios.
\end{enumerate}

\section{Additional remarks on the roles of $C$ and $P$ in the QDM architecture}

In mapping the QDM components onto cognitive functions, it is useful to 
clearly distinguish between the roles of the control register $C$ and the 
policy register $P$. These two subsystems are not redundant but 
complementary, and the contrast between them helps illuminate the sense in 
which the architecture mirrors certain aspects of deliberative cognition.

The control register $C$ may be regarded as the analogue of an executive 
attentional mechanism. At each iteration it specifies which 
branch-dependent unitary $U_b$ is applied to the system register $S$, and 
this choice is then faithfully recorded into a fresh memory slot $M^{(k)}$ 
via the controlled operation $\mathrm{CNOT}_{C\to M^{(k)}}$. In this sense 
$C$ corresponds to the agent’s immediate intention or directive (``what I am 
trying to do now''). One might be tempted to think of $C$ as a merely 
transient selector that disappears once its choice has been deposited in 
memory. Yet the canonical circuits make clear that this is not the case: 
$C$ itself remains present throughout the entire iterative process, 
controlling the system operations in each round and becoming entangled with 
the growing memory string. Thus $C$ is a persistent locus of attentional 
control. Within each branch it is basis-definite, ensuring that the global 
superposition of possibilities remains coherent while still carrying 
definite attentional tags. In cognitive terms, $C$ functions as the 
spotlight of momentary attention, the executive thread that directs what is 
attempted at each step, and that persists as a channel of attentional 
continuity across trials.

The policy register $P$, by contrast, is not an executive selector but the 
analogue of the agent’s ongoing belief state or dispositional background. 
Formally, $P$ is the controlling register for the feedback maps 
$\mathcal{F}^{(k)}$ acting on $S$, and it is itself updated by the 
memory-driven maps $\mathcal{V}^{(k)}$. This dual role gives $P$ a special 
status: it both influences how each trial is internally processed and 
evolves in response to the record of past attentional choices. Unlike $C$, 
the policy register need not be definite within a branch. It may be 
initialized in superposition, become entangled with $S$, or undergo gradual 
reshaping under the influence of memory. In this way $P$ embodies not only 
simple categorical distinctions but also graded, ambiguous, or 
continuously evolving dispositions. It is precisely this adaptability that 
permits us to speak of \emph{memory-driven policy adaptation}: the gradual 
integration of action-history into a persistent dispositional bias.

The temporal ordering of operations makes the functional distinction still 
clearer. In each iteration the circuit applies $\mathcal{U}^{(k)}$ 
(controlled by $C$) to $S$, then records the control in memory, then 
applies feedback $\mathcal{F}^{(k)}$ to $S$ under the influence of $P$, and 
only afterwards applies the update $\mathcal{V}^{(k)}$ to $P$. This ordering 
ensures that $P$ influences the next round of simulation, while being 
reshaped only by the memory of what has already occurred. One may summarize 
the dynamic as follows: $C$ chooses in the short term, $M$ records the fact 
of that choice, and $P$ integrates those records into a longer-term 
dispositional tendency that guides subsequent thinking.

Thus, from a cognitive perspective, the division corresponds to the familiar 
contrast between attention and disposition. The control register $C$ 
embodies the executive channel of attention, not merely a flash at each 
moment but a continuing thread of selection that remains aligned with the 
accumulating diary of past trials. The policy register $P$ embodies the 
slower, more dispositional background: an adaptive substrate of beliefs or 
expectations that persists across time, is continuously reshaped by memory, 
and feeds back into the internal dynamics of the system. Together they 
capture the twofold structure of deliberation: a persistent attentional 
thread directing the present, and a persistent dispositional thread shaping 
the future. 

It should also be noted, however, that the canonical model has a 
limitation. The adaptation of $P$ is conditioned only on the control record 
in $M$, not on the outcome registered in $S$. In effect, the machine learns 
from action-history rather than from error correction (think of a thrill-seeker, 
a fearless explorer, who keeps pushing regardless of the result, solely on 
the basis of the choices previously made). A possible remedy is the 
\emph{overlap-based policy update} (via the swap test 
\cite{Barenco1997,Buhrman2001}), which would extend $P$ to be 
outcome-sensitive, thereby implementing a more genuinely 
self-reflective form of learning. But such modification introduces 
considerable additional structure, and its implications remain 
to be investigated (for a simpler, \emph{outcome-driven, ancilla-mediated} 
approach see Appendix \ref{sec:outcome-oracle-description} and 
Sec.\ \ref{sec:2-iter-example}).

It is useful to consider how the architecture behaves in different 
limiting regimes. If $P$ is initialized in a computational-basis state and each 
$V_b$ maps basis states to basis states, then within each memory-labeled 
branch the policy appears basis-definite (either $\ket{0}_P$ or $\ket{1}_P$), 
and the action of $\mathcal{F}$ is correspondingly unambiguous. At the 
opposite extreme, if $P$ is initialized in superposition, or if the feedback 
maps $F_p$ or the update maps $V_b$ generate superpositions or entanglement, 
then the branch-local state of $P$ need not be basis-definite. Instead it may 
remain in coherent superposition or, when entangled with $S$, in a mixed form,
\begin{equation}
\ket{\Psi}^{(1)} = \sum_{b\in\{0,1\}} \alpha_b \ket{b}_C\ket{b}_M
\Big(\sum_{p\in\{0,1\}} \gamma_p \; F_p U_b\ket{\psi}_S \otimes V_b\ket{p}_P\Big),
\end{equation}
so that the branch-conditional joint $(S,P)$ state is
\begin{equation}
\ket{\Phi_b}_{SP} = \sum_{p}\gamma_p\,F_p U_b\ket{\psi}_S\otimes V_b\ket{p}_P,
\end{equation}
and the branch-conditional reduced policy state is 
$\rho_{P|b}=\operatorname{Tr}_S(\ket{\Phi_b}\bra{\Phi_b})$. These expressions 
underline that $P$ is not simply a classical tag but a genuine quantum register 
whose branch-local form is determined jointly by the initial prior $\ket{\pi}_P$, 
the feedback unitaries $F_p$, the branch dynamics $U_b$, and the updates $V_b$. 
The policy state is therefore reshaped iteratively by the actual path the agent has taken.

Thus, the policy register acquires a twofold interpretation. 
On one hand, it acts as a belief state or disposition, accumulating the 
memory of past control actions and thereby driving policy adaptation through 
$\mathcal{V}$, which in turn biases future simulation through $\mathcal{F}$. 
On the other hand, since $P$ may exist in superposition or be entangled with 
$S$, it can embody ambiguous or blended dispositions, mixtures of competing 
preferences that remain coherent and can interfere when fed back into the 
system. This dual character allows $P$ to represent not just discrete policies 
but also nuanced and context-sensitive dispositions.

The way $P$ is initialized and correlated with $C$ determines markedly different 
behavioral regimes. If $P$ is pre-correlated with $C$ (or initialized to reflect 
recent control history), then feedback acts reflexively within the same iteration, 
producing behavior that closely mirrors the most recent attentional choices. If $P$ 
is initialized independently of $C$ and updated only through $M$, the behavior 
is more deliberative, with $P$ integrating a longer history and biasing future 
rounds without being reactive to the present control decision. Finally, if one 
permits a reflective map $\mathcal{R}^{(k)}:P\to C$ (the reinforcement extension 
of Sec.~\ref{sec:reinforcement}), then the policy state itself can steer the control 
register, altering branch weights and thereby determining which alternatives are 
likely to be entertained in future iterations. This produces yet another cognitive 
regime in which accumulated dispositions directly reshape the space of possibilities 
considered.

From a practical perspective the inclusion of $P$ has several notable consequences. 
It enlarges the representational repertoire of the machine (by permitting persistent, 
graded, or superposed dispositions), it provides an explicit target for interventions 
and for interpretive analysis (since the $\mathcal{V}$-maps determine how policy 
evolves from experience), and it raises resource demands (because maintaining 
coherence in $P$ across multiple iterations is experimentally challenging). 
Whether $P$ is designed to be largely autonomous or tightly coupled to $C$ is 
ultimately a modeling choice, with autonomy favoring deliberative and 
interpretable behavior, and tight coupling favoring reactive and immediate 
responsiveness.

The canonical iteration map
\[
(C \to S)\;\; (C \to M)\;\; (P \to S)\;\; (M \to P)
\]
can thus be interpreted step by step in cognitive terms:
\begin{itemize}
\item[] \textbf{Step 1: $C \to S$ (Attentional Control).}  
The control register $C$ determines which unitary $U_b$ is applied to the 
system $S$. This corresponds to the agent’s momentary focus of attention 
(``let us try option $b$ now''). Within each branch $C$ is definite, so that a 
specific $U_b$ is enacted.

\item[] \textbf{Step 2: $C \to M$ (Memory Encoding).}  
The memory register $M$ records the label generated by $C$, so that each 
branch is tagged with the attentional decision (``I attempted option $b$''). 
The actual outcome of the action is stored in $S$, not in $M$.

\item[] \textbf{Step 3: $P \to S$ (Policy Feedback).}  
The policy register $P$ controls the feedback unitaries $F_p$ that act on $S$. 
This is the agent’s ongoing belief state or disposition, which biases how the 
system evolves after a choice has been made. Unlike $C$, $P$ may be in 
superposition or entangled with $S$, enabling it to encode graded or 
ambiguous beliefs.

\item[] \textbf{Step 4: $M \to P$ (Policy Update / Memory-Driven Adaptation).}  
Once the control decision has been made and recorded, the memory triggers 
an update of $P$ through a unitary $V_b$ controlled by the memory value. 
In this way the fact of which action was attempted gradually reshapes the agent’s 
disposition. This is memory-driven, branch-dependent adaptation based only 
on action records. In the enhanced QDM variant (described in the Appendix), 
a minimal outcome-driven oracle-based policy update mechanism allows 
outcomes in $S$ to contribute as well, thereby enabling a more fully self-reflective 
learning process.
\end{itemize}

\section{Reinforcement Across Branches}
\label{sec:reinforcement}

Up to now we have been mainly concerned with what one may call 
\emph{adaptive} deliberation performed by the QDM. By ``adaptive'' we 
mean a process in which the policy register $P$ is coherently updated 
on the basis of memory-recorded control actions, and this updated policy 
then conditions subsequent feedback operations on the system $S$. 
Adaptive deliberation thus produces history-dependent correlations 
between memories and the policy register, so that the record of past 
memory-encoded choices coherently conditions subsequent feedback on $S$, 
while the relative weights of the main branches, set by the control 
amplitudes $\alpha$ and $\beta$, remain fixed throughout the iterations. 
(Extensions that allow the system state itself to influence policy---for 
example via a simple outcome-driven construction introduced later in the 
Appendix---go 
beyond this strictly canonical form of memory-driven adaptation.)

In contrast, \emph{reinforced} deliberation refers to a related but 
operationally distinct notion, and raises the question of whether 
it can be implemented within the canonical QDM architecture. By 
``reinforcement'' we mean a coherent, history-dependent change in the 
relative weights (probabilities) of alternative evolutions, in which 
past branch-dependent dynamics influence the amplitudes (or outcome 
probabilities) of later alternatives, so that some paths are effectively 
amplified while others are suppressed. In other words, the coefficients 
$\alpha$ and $\beta$ may themselves change between iterations. 
Reinforcement is thus distinct from mere entanglement or bookkeeping of 
histories. It requires a feedback mechanism that re-weights amplitudes 
on the basis of branch history, such that recorded history modifies 
subsequent amplitudes in a way that alters observable outcome 
probabilities.

Two types of reinforcement should be distinguished here:

\vskip5pt

\emph{Subbranch reinforcement.} This is the generic case in the canonical 
architecture with fresh memories $M^{(k)}$. Within a fixed main branch 
(labeled by a memory string $\mathbf{b}$), the policy register $P$ can 
rearrange amplitudes among subbranches, producing constructive or 
destructive interference inside that branch. Because the main-branch 
labels (the memory strings) remain orthogonal, no re-weighting of the 
top-level $C$ amplitudes occurs from intra-branch unitary feedback alone.

\vskip5pt

\emph{Main-branch reinforcement.} Here the aim is to alter the relative 
probabilities of the top-level alternatives ($C=0$ versus $C=1$) under 
unitary evolution. In the canonical architecture, main branches are 
distinguished by distinct memory records. As such, their amplitudes 
remain fixed by the initial superposition of $C$, unless the design of 
the machine is extended in some way.

\vskip5pt

\emph{Possible extension of canonical architecture.} A clean, and arguably 
simplest, modification of our standard architecture may be implemented by 
permitting at each iteration additional coherent reflective feedback from 
the policy register $P$ onto the control register $C$. Operationally this 
means that after the policy update $\mathcal V^{(k)}$ (which uses the 
freshly written memory $M^{(k)}$), we apply a history-dependent unitary 
on $C$ controlled by the updated policy $P$. This leaves all past memory 
records intact (so previous main-branch labels remain well-defined), but 
it changes the state of $C$ before subsequent memory writes, thereby 
biasing which memory strings will be produced in later iterations. In 
this way the policy can coherently ``steer'' the control $C$ and cause 
reinforcement of certain memory-defined main branches in future rounds.

\paragraph{Modified iteration map.}
With that intuitive picture in mind, let us introduce a reflective unitary 
$\mathcal R^{(k)}$ acting on $P\otimes C$, applied after $\mathcal V^{(k)}$. 
The modified $k$-th iteration then has the form,
\begin{equation}
\widetilde{\mathcal W}^{(k)}
\;=\;
\mathcal R^{(k)} \; \mathcal V^{(k)} \; \mathcal F^{(k)} \; 
\mathrm{CNOT}_{C\to M^{(k)}} \; \mathcal U^{(k)}.
\label{eq:Wtilde}
\end{equation}
A convenient and physically transparent choice for $\mathcal R^{(k)}$ is 
a $P$-controlled unitary on $C$,
\begin{equation}
\mathcal R^{(k)} \;=\; \ket{0}\bra{0}_P \otimes R^{(k)}_0
\;+\; \ket{1}\bra{1}_P \otimes R^{(k)}_1,
\end{equation}
where $R^{(k)}_p\in\mathcal B(\mathcal H_C)$ are unitaries on the control 
(qubit), conditioned on the policy outcome $p\in\{0,1\}$. Notice that it is 
conceptually important to apply $\mathcal R^{(k)}$ \emph{after} 
$\mathcal V^{(k)}$, so that it uses the updated policy state for steering 
$C$ before the next iteration.

\paragraph{Illustrative two-iteration example.}
Let us set $\mathcal{U}^{(1)}=\mathcal{U}^{(2)}=\mathbb I$, and ignore 
$\mathcal{F}^{(k)}$, for simplicity. Let us also initialize the machine in
\begin{equation}
\ket{\Psi}^{(0)} = (\alpha\ket{0}_C+\beta\ket{1}_C)\;
\ket{0}_{M^{(1)}}\;\ket{\psi}_S\;\ket{0}_P,
\qquad |\alpha|^2+|\beta|^2=1.
\end{equation}
Next, we choose a simple branch-dependent policy update in the first 
iteration, for example $V^{(1)}_0=\mathbb I_P$ and $V^{(1)}_1=X_P$ (so 
that $P$ becomes $\ket{0}_P$ when $M^{(1)}=0$, and $\ket{1}_P$ when 
$M^{(1)}=1$). After the standard operations up through $\mathcal{V}^{(1)}$ 
(but before $\mathcal{R}^{(1)}$) the state reads,
\begin{equation}
\ket{\Psi}^{(1)} =
\alpha\ket{0}_C\ket{0}_{M^{(1)}}\ket{\psi}_S\ket{0}_P
+
\beta\ket{1}_C\ket{1}_{M^{(1)}}\ket{\psi}_S\ket{1}_P.
\end{equation}

Now apply the reflective map $\mathcal{R}^{(1)}$ with
$R^{(1)}_0=\mathbb I_C$ and $R^{(1)}_1=R_C(\theta)$, where for the control
we take the rotation matrix (in the $\{\ket{0}_C,\ket{1}_C\}$ basis),
\begin{equation}
R_C(\theta)
=
\begin{pmatrix}
\cos\theta & -\sin\theta\\[4pt]
\sin\theta & \cos\theta\end{pmatrix}.
\end{equation}
Acting with $\mathcal{R}^{(1)}$ gives,
\begin{align}
\ket{\Psi_{\rm after\;\mathcal{R}}}^{(1)}
&=
\alpha\ket{0}_C\ket{0}_{M^{(1)}}\ket{\psi}_S\ket{0}_P
+
\beta\Big(\cos\theta\ket{1}_C - \sin\theta\ket{0}_C\Big)
\ket{1}_{M^{(1)}}\ket{\psi}_S\ket{1}_P.
\end{align}

Next, carry out the second iteration's controlled-$U^{(2)}$ (here trivial)
and then write the control into a fresh memory $M^{(2)}$ by
$\mathrm{CNOT}_{C\to M^{(2)}}$. The three terms generated by the state 
above produce three final branches,
\begin{align}
\alpha\ket{0}_C\ket{0}_{M^{(1)}}\ket{0}_{M^{(2)}}\ket{\psi}_S\ket{0}_P
\quad &\longrightarrow\quad \text{memory string } 00 \text{ with amplitude } 
\alpha, \nonumber\\
\beta(-\sin\theta)\ket{0}_C\ket{1}_{M^{(1)}}\ket{0}_{M^{(2)}}\ket{\psi}_S\ket{1}_P
\quad &\longrightarrow\quad \text{memory string } 10 \text{ with amplitude } 
\beta(-\sin\theta), \nonumber\\
\beta\cos\theta\ket{1}_C\ket{1}_{M^{(1)}}\ket{1}_{M^{(2)}}\ket{\psi}_S\ket{1}_P
\quad & \longrightarrow\quad \text{memory string } 11 \text{ with amplitude } 
\beta\cos\theta.
\end{align}
Correspondingly, the final probabilities for the three memory strings are
\begin{equation}
p(00)=|\alpha|^2,
\qquad 
p(10)=|\beta|^2\sin^2\theta,
\qquad 
p(11)=|\beta|^2\cos^2\theta,
\end{equation}
which correctly sum to one (since $|\alpha|^2+|\beta|^2=1$ and 
$\sin^2\theta+\cos^2\theta=1$). Thus, by tuning $\theta$ the reflective 
feedback redistributes the weight among the main-branch memory strings 
emerging after two iterations, and, moreover, brings into play the branch 
labeled $10$, which in the purely canonical architecture would have remained 
unoccupied.

\section{Canonical and outcome-driven QDM architectures}

The preceding sections have introduced the general notion of a quantum 
deliberating machine (QDM) as a closed, self-referential, and fully unitary 
feedback system operating on a finite collection of internal registers. The 
minimal or \emph{canonical} QDM architecture consists of four registers:
\begin{equation}
(C,\, M,\, S,\, P),
\end{equation}
where $C$ is the control register that selects alternative conditional unitaries, 
$M$ is the memory register that records the sequence of control actions taken, 
$S$ is the system under deliberation, and $P$ is the policy register that stores 
a coherent representation of internal decision parameters.  The canonical 
architecture was designed to realize the weakest possible set of requirements 
for self-referential deliberation within a purely quantum, measurement-free 
framework.

In this minimal configuration, the QDM performs what may be called 
\emph{memory-driven deliberation}.  Each deliberation cycle involves 
a reversible mapping
\begin{equation}
C \longrightarrow S \longrightarrow M \longrightarrow P \longrightarrow S,
\end{equation}
where the current state of the system $S$ is influenced by the control $C$, 
its recent action is coherently stored in the memory $M$, and the policy 
register $P$ is updated on the basis of this internal record.  The resulting 
feedback loop allows the machine to evolve in a manner that depends on 
its own action history.  Importantly, no explicit notion of ``success'' or 
``failure'' is invoked at this stage; the QDM simply refines its future dynamics 
by reflecting on its past transformations, thereby implementing a quantum 
form of self-referential adaptation.  This canonical model provides the 
conceptual core of quantum deliberation, stripped of any task-specific 
evaluative mechanism.

However, for many practical or interpretive purposes---such as modeling 
decision processes that involve assessment of outcomes---it is useful to 
enrich this architecture by introducing an ancillary subsystem that explicitly 
encodes performance information.  In such an \emph{outcome-driven} 
or \emph{oracle-augmented} QDM, we add a single ancillary qubit $A$, 
yielding the extended register set
\begin{equation}
(C,\, M,\, S,\, A,\, P).
\end{equation}
Here the ancilla acts as a quantum evaluator or oracle flag. It is entangled 
with the system to indicate whether a given system state satisfies a goal 
predicate.  This ancillary information is then used to condition the update 
of the policy register $P$, typically via an anti-CNOT or controlled-phase 
operation.  The modified policy subsequently feeds back into the system 
$S$ during the next iteration, producing outcome-dependent adaptation.

The canonical architecture and the outcome-driven extension thus represent 
two levels of deliberative capability:
\begin{enumerate}
\item \textbf{Canonical QDM (memory-driven deliberation):} The policy evolves 
solely from the coherent record of prior actions, without any external or 
oracle-based evaluation.  This form of deliberation captures the intrinsic 
capacity of the machine to refer to its own history.
\item \textbf{Outcome-driven QDM (ancilla-mediated deliberation):} The 
policy is updated not only from action history but also from internalized 
performance feedback provided by the ancilla register.  This permits adaptive 
refinement toward specific goal states.
\end{enumerate}

The canonical case serves as the foundational construction of the QDM 
concept: it embodies the logical structure of self-referential, 
amplitude-preserving deliberation.  The introduction of the ancilla $A$ 
in the subsequent example does not modify this foundation but rather 
extends it to include the simplest possible evaluative channel.  The 
two-iteration example presented below (Section \ref{sec:2-iter-example}) 
illustrates explicitly how such an outcome-driven feedback cycle operates, 
demonstrating how a QDM can achieve coherent convergence toward a 
target state while preserving full unitarity.

\section{Motivation for the Outcome-Driven Two-Iteration Example}

Before presenting the explicit two-iteration construction 
(Sec.\ \ref{sec:2-iter-example}), it is useful to 
clarify its role and level of abstraction. Although the example represents 
an extremely simple deliberative process (essentially a single corrective 
feedback step), it serves as the minimal working demonstration of 
outcome-sensitive, self-referential evolution within the QDM framework. 
In the canonical $(C,M,S,P)$ architecture, policy adaptation is memory-driven, 
depending only on which actions were taken. The present extension 
introduces an additional ancilla register $A$, allowing the policy update to 
depend explicitly on the observed outcome of the previous iteration. This 
makes the process outcome-driven, albeit in the simplest possible sense. 

Despite its apparent triviality from a computational standpoint, the two-iteration 
example is conceptually important. It provides the smallest closed deliberation 
loop in which all essential mechanisms (controlled branching of the system 
by $C$, coherent recording of branch history in $M$, policy feedback and update 
through $P$, and optional outcome mediation via $A$) are explicitly realized. 
The simplicity of the setup allows one to track the full entangled dynamics across 
all registers and to examine how coherent feedback and interference among 
policy branches can stabilize or redirect the system’s evolution. 

The same structural pattern can be straightforwardly generalized to 
multi-iteration or continuous-feedback architectures. In an $n$-iteration QDM, 
the policy register $P$ becomes a dynamically evolving quantum memory, 
coherently integrating information from successive $M$ and $A$ updates, 
thus encoding a superposition of past deliberative trajectories. Each iteration 
can be viewed as a unitary ``policy update gate'' acting on $(P,M,A)$, conditionally 
modifying the system’s future behavior. In the continuous-feedback limit, the 
discrete policy updates would be replaced by a policy Hamiltonian generating gradual 
adaptation of $P$ in real time, effectively realizing a form of coherent learning 
dynamics. These generalized forms would preserve the essential feature first exhibited 
in the two-iteration model: the internal self-referential loop in which control, 
memory, system, and policy interact coherently to maintain, reinforce, or 
redirect the system’s state through interference-based stabilization and 
adaptive policy evolution.

\section{Application: Coherent Search-and-Rescue}
\label{sec:2-iter-example}

In this section we present a simple 2-iteration example in which 
outcome-driven oracle-based adaptive policy update ``helps'' the system 
$S$ arrive at a given target state (here chosen to be $\ket{1}_S$), even 
though initially (in first iteration) the control $C$ applied two \emph{different} 
branch-dependent unitaries to $S$, forcing the system $S$ to initially evolve 
in some arbitrarily chosen ``different directions.'' This is the simplest nontrivial 
case where policy update genuinely changes later evolution. The example 
makes the abstract ``deliberation'' process fully explicit, showing step-by-step 
register entanglement and adaptive feedback. It thus answers the ``What actually 
happens inside the QDM?'' question.

\subsection{Main construction}

The full machine has the 5 standard canonical registers, $(C,M^{(1)},M^{(2)},S,P)$, 
and an additional ancilla register, $A$, initialized in $\ket{0}_A$, which facilitates 
outcome-driven policy adaptation (see Appendix \ref{sec:outcome-oracle-description}). 
The full register lineup is $(C,M^{(1)},M^{(2)},S,A,P)$. The machine is initialized 
in the tensor-product state,
\begin{equation}
\ket{\Psi}^{(0)}
=
(\alpha \ket{0}_C + \beta \ket{1}_C)
\ket{0}_{M^{(1)}}\ket{0}_{M^{(2)}}
\ket{0}_{S}
\ket{0}_{A}
\ket{0}_{P}.
\end{equation}

The sequence of operations and the resulting QDM states in the first iteration are: 
\begin{enumerate}
\item $C\to S$,
\begin{align}
\mathcal{U}&=\ket{0}\bra{0}_C \otimes U_{0} + \ket{1}\bra{1}_C \otimes U_{1}, 
\\
\ket{\Psi} 
&=
[\alpha \ket{0}_C \ket{0}_{M^{(1)}}\ket{0}_{M^{(2)}} (U_{0,00} \ket{0}_{S} + U_{0,10} \ket{1}_{S}) 
\nonumber \\
& \quad
+ \beta \ket{1}_C \ket{0}_{M^{(1)}}\ket{0}_{M^{(2)}} (U_{1,00} \ket{0}_{S} + U_{1,10} \ket{1}_{S}) ]
\ket{0}_{A}
\ket{0}_{P} ,
\end{align}
\item $C\to M^{(1)}$, 
\begin{align}
\text{CNOT}_{C\to M^{(1)}}&=\ket{0}\bra{0}_C \otimes I_{M^{(1)}} + \ket{1}\bra{1}_C \otimes X_{M^{(1)}}, 
\\
\ket{\Psi} 
&=
[\alpha \ket{0}_C \ket{0}_{M^{(1)}}\ket{0}_{M^{(2)}} (U_{0,00} \ket{0}_{S} + U_{0,10} \ket{1}_{S}) 
\nonumber \\
& \quad
+ \beta \ket{1}_C \ket{1}_{M^{(1)}}\ket{0}_{M^{(2)}} (U_{1,00} \ket{0}_{S} + U_{1,10} \ket{1}_{S}) ]
\ket{0}_{A}
\ket{0}_{P} ,
\end{align}
\item $P\to S$ (policy-driven feedback, fixed by design), 
\begin{align}
\mathcal{F}&=\text{CNOT}_{P\to S} = \ket{0}\bra{0}_P \otimes I_{S} + \ket{1}\bra{1}_P \otimes X_{S}, 
\\
\ket{\Psi} 
&=
[\alpha \ket{0}_C \ket{0}_{M^{(1)}}\ket{0}_{M^{(2)}} 
(U_{0,00} \ket{0}_{S} + U_{0,10} \ket{1}_{S} ) 
\nonumber \\
& \quad
+ \beta \ket{1}_C \ket{1}_{M^{(1)}}\ket{0}_{M^{(2)}} 
(U_{1,00} \ket{0}_{S} + U_{1,10} \ket{1}_{S}) ] \ket{0}_{A}\ket{0}_{P} 
\qquad \text{(no change of state)},
\end{align}
\item $S\to A$ (outcome-evaluation oracle used to ``flag'' the ancilla), 
\begin{align}
\mathcal{O}&=\text{CNOT}_{S\to A} =\ket{0}\bra{0}_S \otimes I_{A} + \ket{1}\bra{1}_S \otimes X_{A}, 
\\
\ket{\Psi} 
&=
\alpha \ket{0}_C \ket{0}_{M^{(1)}}\ket{0}_{M^{(2)}} 
(U_{0,00} \ket{0}_{S}\ket{0}_{A}\ket{0}_{P} + U_{0,10} \ket{1}_{S}\ket{1}_{A}\ket{0}_{P} ) 
\nonumber \\
& \quad
+ \beta \ket{1}_C \ket{1}_{M^{(1)}}\ket{0}_{M^{(2)}} 
(U_{1,00} \ket{0}_{S} \ket{0}_{A}\ket{0}_{P} + U_{1,10} \ket{1}_{S}\ket{1}_{A}\ket{0}_{P} ) ,
\end{align}
\item $A\to P$ (ancilla-controlled policy update; if the ``good'' system state, $\ket{1}_S$, 
is detected, do nothing, if the ``bad'' system state, $\ket{0}_S$, is detected, flip the policy), 
\begin{align}
\text{antiCNOT}_{A\to P}&=\ket{0}\bra{0}_A \otimes X_{P} + \ket{1}\bra{1}_A \otimes I_{P}, 
\\
\ket{\Psi} 
&=
\alpha \ket{0}_C \ket{0}_{M^{(1)}}\ket{0}_{M^{(2)}} 
(U_{0,00} \ket{0}_{S}\ket{0}_{A}\ket{1}_{P} + U_{0,10} \ket{1}_{S}\ket{1}_{A}\ket{0}_{P} ) 
\nonumber \\
& \quad
+ \beta \ket{1}_C \ket{1}_{M^{(1)}}\ket{0}_{M^{(2)}} 
(U_{1,00} \ket{0}_{S} \ket{0}_{A}\ket{1}_{P} + U_{1,10} \ket{1}_{S}\ket{1}_{A}\ket{0}_{P} ) ,
\end{align}
\item $S\to A$ (inverse oracle used to uncompute the ancilla), 
\begin{align}
\mathcal{O}^{\dagger}&=\text{CNOT}_{S\to A} =\ket{0}\bra{0}_S \otimes I_{A} + \ket{1}\bra{1}_S \otimes X_{A}.
\\
\ket{\Psi} 
&=
\alpha \ket{0}_C \ket{0}_{M^{(1)}}\ket{0}_{M^{(2)}} 
(U_{0,00} \ket{0}_{S}\ket{0}_{A}\ket{1}_{P} + U_{0,10} \ket{1}_{S}\ket{0}_{A}\ket{0}_{P} ) 
\nonumber \\
& \quad
+ \beta \ket{1}_C \ket{1}_{M^{(1)}}\ket{0}_{M^{(2)}} 
(U_{1,00} \ket{0}_{S} \ket{0}_{A}\ket{1}_{P} + U_{1,10} \ket{1}_{S}\ket{0}_{A}\ket{0}_{P} ) .
\end{align}
\end{enumerate}

The sequence of operations and the resulting QDM states in the second iteration is: 
\begin{enumerate}
\item $C\to S$,
\begin{align}
\mathcal{U}&=I_{CS}  \qquad \text{(identity, fixed by design)}, 
\\
\ket{\Psi} 
&=
\alpha \ket{0}_C \ket{0}_{M^{(1)}}\ket{0}_{M^{(2)}} 
(U_{0,00} \ket{0}_{S}\ket{0}_{A}\ket{1}_{P} + U_{0,10} \ket{1}_{S}\ket{0}_{A}\ket{0}_{P} ) 
\nonumber \\
& \quad
+ \beta \ket{1}_C \ket{1}_{M^{(1)}}\ket{0}_{M^{(2)}} 
(U_{1,00} \ket{0}_{S} \ket{0}_{A}\ket{1}_{P} + U_{1,10} \ket{1}_{S}\ket{0}_{A}\ket{0}_{P} ) 
\end{align}
\item $C\to M^{(2)}$,
\begin{align}
\text{CNOT}_{C\to M^{(2)}}&=\ket{0}\bra{0}_C \otimes I_{M^{(2)}} + \ket{1}\bra{1}_C \otimes X_{M^{(2)}}, 
\\
\ket{\Psi} 
&=
\alpha \ket{0}_C \ket{0}_{M^{(1)}}\ket{0}_{M^{(2)}} 
(U_{0,00} \ket{0}_{S}\ket{0}_{A}\ket{1}_{P} + U_{0,10} \ket{1}_{S}\ket{0}_{A}\ket{0}_{P} ) 
\nonumber \\
& \quad
+ \beta \ket{1}_C \ket{1}_{M^{(1)}}\ket{1}_{M^{(2)}} 
(U_{1,00} \ket{0}_{S} \ket{0}_{A}\ket{1}_{P} + U_{1,10} \ket{1}_{S}\ket{0}_{A}\ket{0}_{P} ) 
\end{align}
\item $P\to S$ (policy-driven feedback, fixed by design), 
\begin{align}
\mathcal{F}&=\text{CNOT}_{P\to S} =\ket{0}\bra{0}_P \otimes I_{S} + \ket{1}\bra{1}_P \otimes X_{S}, 
\\
\ket{\Psi} 
&=
\alpha \ket{0}_C \ket{0}_{M^{(1)}}\ket{0}_{M^{(2)}} 
\ket{1}_{S}\ket{0}_{A} (U_{0,00} \ket{1}_{P} + U_{0,10} \ket{0}_{P} ) 
\nonumber \\
& \quad
+ \beta \ket{1}_C \ket{1}_{M^{(1)}}\ket{1}_{M^{(2)}} 
\ket{1}_{S} \ket{0}_{A}(U_{1,00} \ket{1}_{P} + U_{1,10} \ket{0}_{P} ) ,
\end{align}
at which point, \emph{in both branches}, the system arrived at the desired 
state $\ket{1}_S$, and, in principle, the QDM may now stop (the goal has been achieved!). 
Nevertheless, it is useful to formally complete the second iteration by verifying 
that the additional oracle-based evaluation will not change this state, so we 
proceed in the same way as in first iteration.
\item $S\to A$ (outcome-evaluation oracle used to ``flag'' the ancilla), 
\begin{align}
\mathcal{O}&=\text{CNOT}_{S\to A} =\ket{0}\bra{0}_S \otimes I_{A} + \ket{1}\bra{1}_S \otimes X_{A}, 
\\
\ket{\Psi} 
&=
\alpha \ket{0}_C \ket{0}_{M^{(1)}}\ket{0}_{M^{(2)}} 
\ket{1}_{S}\ket{1}_{A} (U_{0,00} \ket{1}_{P} + U_{0,10} \ket{0}_{P} ) 
\nonumber \\
& \quad
+ \beta \ket{1}_C \ket{1}_{M^{(1)}}\ket{1}_{M^{(2)}} 
\ket{1}_{S} \ket{1}_{A}(U_{1,00} \ket{1}_{P} + U_{1,10} \ket{0}_{P} ) ,
\end{align}
\item $A\to P$ (ancilla-controlled policy update; if the ``good'' system state, 
$\ket{1}_S$, is detected, do nothing, if the ``bad'' system state, $\ket{0}_S$, 
is detected, flip the policy), 
\begin{align}
\text{antiCNOT}_{A\to P}&=\ket{0}\bra{0}_A \otimes X_{P} + \ket{1}\bra{1}_A \otimes I_{P}, 
\\
\ket{\Psi} 
&=
\alpha \ket{0}_C \ket{0}_{M^{(1)}}\ket{0}_{M^{(2)}} 
\ket{1}_{S}\ket{1}_{A} (U_{0,00} \ket{1}_{P} + U_{0,10} \ket{0}_{P} ) 
\nonumber \\
& \quad
+ \beta \ket{1}_C \ket{1}_{M^{(1)}}\ket{1}_{M^{(2)}} 
\ket{1}_{S} \ket{1}_{A}(U_{1,00} \ket{1}_{P} + U_{1,10} \ket{0}_{P} ) ,
\end{align}
\item $S\to A$ (inverse oracle used to uncompute the ancilla), 
\begin{align}
\mathcal{O}^{\dagger}&=\text{CNOT}_{S\to A} =\ket{0}\bra{0}_S 
\otimes I_{A} + \ket{1}\bra{1}_S \otimes X_{A}.
\\
\ket{\Psi} 
&=
\alpha \ket{0}_C \ket{0}_{M^{(1)}}\ket{0}_{M^{(2)}} 
\ket{1}_{S}\ket{0}_{A} (U_{0,00} \ket{1}_{P} + U_{0,10} \ket{0}_{P} ) 
\nonumber \\
& \quad
+ \beta \ket{1}_C \ket{1}_{M^{(1)}}\ket{1}_{M^{(2)}} 
\ket{1}_{S} \ket{0}_{A}(U_{1,00} \ket{1}_{P} + U_{1,10} \ket{0}_{P} ) ,
\end{align}
as expected. We note that while in both branches the system ended up 
in the target state, the policy ended up in a branch-dependent superposition 
of 0 and 1 (the quantum analogue of uncertain but coherent policy weights), 
with coefficients determined by the matrix elements of the unitary operators 
applied to the system register.
\end{enumerate}

A note on resources: with 6 qubits and 2 iterations, we ended up needing 
12 two-qubit gates, which corresponds to scaling $\propto$ number of 
iterations $\times$ number of registers, illustrating linear growth with 
iteration count.

The corresponding circuit is shown in Fig.~\ref{fig:2-iter-enhanced_outcome}. 
This minimal two-iteration circuit explicitly demonstrates that outcome-driven 
policy updates can bias later evolution without measurements and coherently 
steer the system toward a target state. After the first iteration, the policy qubit 
becomes entangled with the system's outcome, encoding whether the 
corresponding branch achieved the target. In the second iteration, this 
policy information coherently drives both branches of the control register 
so that the system deterministically reaches the desired $\ket{1}_S$ state. 
\begin{figure}[H]
\centering
\scalebox{0.9}{\begin{quantikz}[row sep=0.5cm, column sep=0.6cm]
\lstick{$C$}            & \gate[wires=1]{\alpha\ket{0}+\beta\ket{1}} & \ctrl{3} & \ctrl{1} & \qw& \qw& \qw & \qw  
                             & \qw      & \ctrl{2} & \qw& \qw& \qw & \qw  \\
\lstick{$M^{(1)}$} & \qw      & \qw      & \targ{} & \qw& \qw& \qw & \qw  
                             & \qw      & \qw & \qw& \qw& \qw & \qw  \\
\lstick{$M^{(2)}$} & \qw      & \qw       & \qw & \qw& \qw& \qw & \qw  
                             & \qw       & \targ{} & \qw& \qw& \qw & \qw  \\
\lstick{$S$} & \qw      & \gate[wires=1]{U_0/U_1} & \gate[wires=1]{I/X} &\ctrl{1} & \qw& \ctrl{1} & \qw 
                   & \qw                                                   & \gate[wires=1]{I/X} &\ctrl{1} & \qw& \ctrl{1} & \qw  \\
\lstick{$A$} & \qw      & \qw & \qw & \targ{} & \octrl{1} & \targ{} & \qw 
                   & \qw & \qw & \targ{} & \octrl{1} & \targ{} & \qw  \\
\lstick{$P$} & \qw      & \qw & \ctrl{-2} & \qw& \targ{} & \qw & \qw 
                   & \qw & \ctrl{-2} & \qw& \targ{} & \qw & \qw 
\end{quantikz}
}
\caption{Quantum circuit for the 2-iteration QDM that implements 
outcome-driven oracle-based adaptive policy update, which helps steer 
the system register $S$
to the desired target state $\ket{1}_S$, regardless of the initial branch-dependent 
``push'' performed by control $C$.}
\label{fig:2-iter-enhanced_outcome}
\end{figure}

\subsection{Remarks}

\paragraph*{Policy superposition and its significance.}
After the second deliberation step, the system register $S$ deterministically 
reaches the target state $\ket{1}_S$, while the policy qubit $P$ no longer 
assumes a definite classical value. Instead, it remains in a branch--dependent 
superposition of 
$\ket{0}_P$ and $\ket{1}_P$ whose amplitudes depend explicitly on the 
complex transition coefficients of the branch unitaries $U_0$ and $U_1$.  
This \emph{policy superposition} encodes a coherent record of the system’s 
response amplitudes rather than of its measured outcomes, thereby 
representing a fully quantum form of internal adaptation.  In contrast to 
classical reinforcement learning, where a single stochastic policy is updated 
according to scalar rewards, the QDM retains amplitude-level information 
about its own success structure, enabling interference between alternative 
“strategies” in subsequent iterations.  The policy register thus embodies a 
\emph{coherent mixture of learned rules}, entangled with the system’s 
deliberative history.  Such entanglement allows later updates to exploit 
constructive or destructive interference between policies, realizing a form 
of amplitude-based credit assignment with no classical analogue.  The 
existence of this branch-dependent policy superposition is the defining 
operational signature of a genuinely quantum deliberation process---absent 
in any classical learning system---and may represent a new primitive for 
adaptive quantum algorithms and meta-learning architectures.

\paragraph*{Policy-induced backaction.}
An interesting consequence of the outcome-driven adaptive cycle is that 
the ``learned'' policy qubit $P$, once left in a coherent superposition 
determined by the branch unitaries $U_i$, will in the next deliberation 
step coherently apply both $I_S$ and $X_S$ to the system.  As a result, 
the system is driven \emph{out} of the previously achieved target state, 
reintroducing quantum uncertainty ({\it cf}.\ Grover's search algorithm). 
This apparent ``instability'' (loss of target) is not merely a flaw, 
but a structural signature of coherent policy encoding. 
The machine has not classically committed to a single 
policy, but instead retains a superposed internal representation of 
alternative control strategies. In this sense, the backaction of the policy 
superposition on the system marks the transition from classical convergence 
to genuinely quantum, interference-enabled re-deliberation. 
Whether such coherence should be preserved, suppressed, or exploited 
depends on the intended operational regime of the QDM---stable target 
attainment versus exploratory quantum adaptation.

\paragraph*{Contrast with the canonical architecture.}
It is instructive to compare this outcome-driven feedback cycle with the 
canonical, ancilla-free QDM architecture discussed earlier. In the canonical 
case, the policy register $P$ evolves solely under the influence of the internal 
memory record $M^{(k)}$, without any oracle-like signal indicating whether 
a given system configuration was favorable or not.  The resulting adaptation 
is therefore purely self-referential; it encodes structural correlations among 
past actions, but lacks any amplitude bias toward a designated goal state.  
Under such memory-driven deliberation, the evolution of $S$ remains 
consistent with the machine’s internal history but cannot autonomously 
discriminate between successful and unsuccessful branches.

By contrast, in the present outcome-driven version, the ancillary qubit $A$ 
introduces an explicit evaluative channel that allows amplitude-level biasing 
of the policy.  The ancilla’s conditional interaction with the policy register 
converts information about outcome quality (encoded as quantum 
correlations) into a coherent modification of the internal policy state.  
This modification then acts back on $S$ in the next iteration, coherently 
steering it toward the target subspace.  The two-iteration example thus 
demonstrates a qualitatively new mode of deliberation: outcome-sensitive 
policy feedback that remains fully unitary and reversible. In this sense, the 
addition of $A$ transforms the QDM from a self-referential but neutral 
deliberator into an internally adaptive agent capable of coherent 
goal-directed refinement.

\paragraph*{On the use of an ancilla for policy updates.}
In the example above we implemented outcome-driven policy adaptation 
via an ancilla register $A$ (compute-act-uncompute pattern). A superficially 
simpler implementation would replace the three-gate sequence 
$S\!\to\!A\to P\to S$ by a single controlled update $S\!\to\!P$ 
(antiCNOT controlled by $S$). While the direct update is cheaper in qubits 
and gates, it is not equivalent in scope. The ancilla pattern permits 
reversible computation of arbitrary Boolean conditions (including functions 
of the memory registers $M^{(i)}$), selective flipping of 
$P$, and clean uncomputation of workspace so that no garbage accumulates 
across iterations. Moreover, although ancilla mediation resets the workspace, 
it does not in general avoid entanglement between $S$ and $P$ if $S$ is in 
superposition --- an unavoidable consequence of coherent conditional updating.  
Thus, for minimal toy demonstrations where $S$ is in a basis state or where 
S--P entanglement is acceptable, the direct $S\!\to\!P$ gate is an efficient choice; 
for the full QDM architecture (where complex branch-dependent logic and 
reversible bookkeeping are required) the ancilla-mediated compute--act--uncompute 
pattern is the more appropriate primitive.

\paragraph*{Dependence on initial policy state.}
We note the following important caveat of the two-iteration illustration above.
The deterministic steering of $S$ to $\ket{1}_S$ in both branches relies 
on the policy qubit being initialized in the definite state $\ket{0}_P$.  
If $P$ is instead prepared in a superposition $\gamma\ket{0}_P+\delta\ket{1}_P$, 
the conditional updates entangle $S$ and $P$ and the second-iteration 
feedback no longer guarantees that $S$ collapses to $\ket{1}_S$ in both 
branches. Thus, the toy protocol is not ``universal'' with respect to arbitrary 
prior policy states. This is not a fundamental refutation of the QDM paradigm, 
but it does impose a design constraint. One must either (i) supply 
a policy-initialization/reset stage (measurement or dissipative reset), 
(ii) implement a coherent policy-preparation subroutine prior to 
deliberation, or (iii) employ multi-step interference/amplitude-amplification 
strategies that dilute the influence of an unknown initial policy over several 
iterations.  

\paragraph*{Classical analogue and conceptual advantage.}
The classical analogue of the two-iteration outcome-driven QDM is a 
finite-state feedback controller that records its recent actions, evaluates 
success via a binary flag, and deterministically updates an internal policy 
variable. Such an automaton operates serially, updating a definite policy 
state after each evaluation. In contrast, the QDM performs these updates 
coherently across all control branches. Its policy register becomes a 
superposition of alternative strategies, and subsequent system evolution 
exploits quantum interference to favor successful amplitudes. Although 
this minimal two-iteration construction does not confer any computational 
speedup over its classical counterpart, it demonstrates a qualitatively new 
property, namely, the possibility of outcome-dependent adaptation within 
a fully unitary and reversible dynamics. This establishes a foundational 
distinction between classical feedback learning and quantum deliberation.
The latter can refine its internal policy without collapsing superposition or 
discarding past amplitude information.

\paragraph*{QDM as a quantum autopilot.}
Thus, from a control-theoretic perspective, the outcome-driven QDM realizes 
the minimal architecture of a quantum autopilot. The control register 
$C$ acts as the onboard controller selecting an action, the system $S$ 
as the physical device undergoing the action, the ancilla $A$ as the sensor 
or evaluator of performance, the memory $M$ as the internal flight log, 
and the policy register $P$ as the adaptive control law. 
Their closed-loop interaction implements a self-correcting 
cycle in which the machine autonomously refines its control 
policy based on internally generated outcome information. 
In this sense, the two-iteration construction embodies the smallest 
coherent feedback controller---a quantum autopilot that learns to 
steer its own dynamics toward a goal state without external 
supervision or measurement-induced collapse.

\paragraph*{QDM as quantum search and rescue.}
The two-iteration outcome-driven QDM may also be viewed metaphorically 
as a quantum search-and-rescue device.  In this picture, the control register 
$C$ launches the system $S$ along alternative trajectories (different 
``wanderers'' exploring a landscape). Some trajectories, determined by the 
$C$-controlled unitaries $U_i$, begin far from the target region, while 
others may approach it by chance.  The ancilla $A$ acts as a coherent 
locator beacon, flagging which branches are near the desired goal state. 
Through the feedback chain $A \rightarrow P \rightarrow S$, the internal 
policy updates its control law so that, in the next iteration, amplitude flows 
preferentially toward the ``rescued'' branches.  The resulting constructive 
interference of successful paths and attenuation of unsuccessful ones mirrors 
the amplitude amplification mechanism familiar from Grover-type quantum 
search, but here it arises from an internally generated, deliberative feedback 
loop. In this sense, the QDM behaves as a quantum search-and-rescue 
autopilot, capable of steering itself toward a goal regardless of how its 
initial branches were distributed.

\section{Two Machines in Conversation}

In the canonical model, a single quantum deliberating machine (QDM) 
evolves by iterating an internal update map that 
acts on its control ($C$), system ($S$), memory ($M$), and policy ($P$) registers.
Each iteration $k$ consists of the following temporally ordered steps:
\begin{enumerate}
    \item Apply a branch-dependent unitary $\mathcal{U}^{(k)}$ to $C \otimes S$.
    \item Copy the control bit into the fresh memory register via 
    $\mathrm{CNOT}_{C \to M^{(k)}}$.
    \item Apply a feedback unitary $\mathcal{F}^{(k)}$ from $P$ to $S$.
    \item Apply a reflective update $\mathcal{V}^{(k)}$ from $M^{(k)}$ to $P$.
\end{enumerate}

Consider now two such machines, QDM-A and QDM-B, with registers
\[
\mathcal{H}_A = \mathcal{H}_{C_A} \otimes \mathcal{H}_{M_A} 
\otimes \mathcal{H}_{S_A} \otimes \mathcal{H}_{P_A}, \quad
\mathcal{H}_B = \mathcal{H}_{C_B} \otimes \mathcal{H}_{M_B} 
\otimes \mathcal{H}_{S_B} \otimes \mathcal{H}_{P_B}.
\]
The joint Hilbert space is $\mathcal{H}_{AB} = \mathcal{H}_A \otimes \mathcal{H}_B$.
In addition to their internal deliberative cycles, the two machines may 
exchange information through a \emph{conversation channel}.
We distinguish three basic types of such channels, based on which 
architecture layer is coupled:

\begin{itemize}
    \item[(a)] \textbf{System-to-system coupling.} 
    A unitary interaction acts on $S_A \otimes S_B$, for example 
    $\exp(i \theta Z_{S_A} \otimes Z_{S_B})$.,
\begin{equation}
\mathcal{X}^{(k)}_{AB}
=
I_{C_A M_A^{(k)} P_A}
\otimes \exp(i \theta Z_{S_A} \otimes Z_{S_B}) 
\otimes I_{C_B M_B^{(k)} P_B}
\end{equation}
    Cognitively, this represents the exchange of ``utterances'' or 
    surface-level proposals between the two machines.

    \item[(b)] \textbf{Policy-to-policy coupling.} 
    One policy register serves as a control for an operation on 
    the other machine. 
    For instance,
    \begin{equation}
    \mathcal{X}^{(k)}_{AB} = 
    I_{C_A M_A^{(k)} S_A}
    \otimes 
    \left( \ket{0}\bra{0}_{P_A} \otimes I_B 
    + 
    \ket{1}\bra{1}_{P_A} \otimes H^{(k)}_B \right),
    \end{equation}
    meaning that QDM-A’s policy biases QDM-B’s deliberation. 
    Cognitively, this is analogous to persuasion or argumentation, 
    where one agent’s stance conditions the other’s processing.

    \item[(c)] \textbf{Memory exchange.} 
    One machine’s freshly written memory register is transferred into 
    the other’s memory bank. 
    Cognitively, this corresponds to recounting past experience.
\end{itemize}

Let $\mathcal{W}_A^{(k)}$ denote the internal update of QDM-A at iteration $k$,
and similarly $\mathcal{W}_B^{(k)}$ for QDM-B.
Each is defined in terms of the temporally ordered operators
\[
\mathcal{W}_A^{(k)} = 
\mathcal{V}_A^{(k)} \,
\mathcal{F}_A^{(k)} \,
\mathrm{CNOT}_{C_A \to M_A^{(k)}} \,
\mathcal{U}_A^{(k)} ,
\]
with each acting nontrivially only on its designated registers and as identity elsewhere.
We define the global iteration operator as
\[
\mathcal{W}^{(k)}_{AB} =
\begin{cases}
\mathcal{X}^{(k)}_{AB}
\big( \mathcal{W}_A^{(k)} \otimes I_B \big)
\big( I_A \otimes \mathcal{W}_B^{(k)} \big), & \text{(deliberate first, then converse)}, \\[6pt]
\big( \mathcal{W}_A^{(k)} \otimes I_B \big)
\big( I_A \otimes \mathcal{W}_B^{(k)} \big)
\mathcal{X}^{(k)}_{AB}, & \text{(converse first, then deliberate)}.
\end{cases}
\]
Here $I_A$ (resp.\ $I_B$) denotes the identity on all registers of QDM-A (resp.\ QDM-B).
The choice of temporal ordering leads to distinct dynamics:
in the first case (Fig.\ \ref{fig:delib-first}), each machine completes its private round of deliberation 
before exchanging utterances; in the second case, the conversation occurs 
at the start of the round and thereby shapes the subsequent internal loop.

This construction preserves the defining features of a single QDM 
(self-referential deliberation through reflective feedback), 
while adding an external dialogue channel.
Depending on the nature of $\mathcal{X}^{(k)}_{AB}$, 
the machines may be understood as exchanging immediate outputs (system-to-system), 
influencing each other's dispositions (policy-to-policy), 
or recounting remembered experiences (memory exchange).
The conversation is itself coherent and branch-dependent, 
so that parallel, entangled deliberations can unfold across the two machines.
\begin{figure}[H]
\centering
\begin{quantikz}[row sep=0.5cm, column sep=0.6cm]
\lstick{$C_A$}           & \gate{\alpha\ket{0}+\beta\ket{1}}     & \ctrl{2}               & \ctrl{1}                & \qw                     & \qw & \qw \\
\lstick{$M_A^{(k)}$} & \qw                                                    & \qw                    & \targ{}                & \ctrl{2}                 & \qw & \qw \\
\lstick{$S_A$}           & \qw                                                    & \gate{U_A^{(k)}} & \gate{F_A^{(k)}} & \qw                     & \ctrl{4} & \qw \\
\lstick{$P_A$}           & \gate{\gamma\ket{0}+\delta\ket{1}} & \qw                    & \ctrl{-1}               & \gate{V_A^{(k)}} & \qw & \qw \\
\lstick{$C_B$}           & \gate{\alpha'\ket{0}+\beta'\ket{1}}    & \ctrl{2}               & \ctrl{1}                & \qw                     & \qw & \qw \\
\lstick{$M_B^{(k)}$} & \qw                                                    & \qw                    & \targ{}                 & \ctrl{2}                & \qw & \qw \\
\lstick{$S_B$}           & \qw                                                    & \gate{U_B^{(k)}} & \gate{F_B^{(k)}} & \qw                      & \targ{} & \qw \\
\lstick{$P_B$}          & \gate{\gamma'\ket{0}+\delta'\ket{1}} & \qw                    & \ctrl{-1}              & \gate{V_B^{(k)}}  & \qw & \qw
\end{quantikz}
\caption{Deliberate first, then converse. Each QDM executes its internal update cycle
$ \mathcal{U}^{(k)}$, $ \mathrm{CNOT}_{C\to M^{(k)}}$, $ \mathcal{F}^{(k)}$, $ \mathcal{V}^{(k)}$. After the internal operations, the
two system registers $S_A$ and $S_B$ interact via the conversation gate 
$G^{(k)}_{AB}$, schematically drawn here as a control on $S_A$ and a 
target on $S_B$.}
\label{fig:delib-first}
\end{figure}

\section{Higher-Order and Categorical Perspective on the QDM}

The QDM can also be viewed through the lens of higher-order quantum 
processes. In standard quantum mechanics, a unitary acts on a quantum 
state, producing a new state. In contrast, higher-order processes act 
on the unitaries themselves, coherently selecting, modulating, or 
composing them based on the state of ancillary registers. In the 
QDM, the control register $C$ determines which branch-dependent 
unitary is applied to the system $S$, while the memory registers 
$M^{(k)}$ carry records of past control choices that can influence 
subsequent system dynamics. Simultaneously, the policy register $P$ 
encodes adaptive preferences that bias or adjust future operations. 
Together, these registers implement a form of coherent, 
branch-dependent modulation of system unitaries, realizing a higher-order map 
in which quantum states effectively govern the evolution of other quantum 
operations (cf.~\cite{Chiribella2008, Chiribella2009, Bera2019}).

From the categorical standpoint, higher-order processes are naturally 
described as morphisms that take processes as inputs and yield new 
processes as outputs. The QDM fits naturally into this framework. 
The evolution of the system register $S$ is a process whose structure is 
dynamically shaped by the states of $C$, $M$, and $P$. This positions 
the QDM alongside constructions such as supermaps and the quantum 
switch \cite{Chiribella2013,Oreshkov2012,Hardy2007}, which provide 
abstract, compositional descriptions of processes acting on processes. 
Frameworks for higher-order processes with non-fixed causal structure 
provide an abstract basis for understanding how the 
QDM’s branch-dependent, feedback-modulated operations can be 
coherently composed, generalizing beyond standard fixed-order circuits.
By framing the QDM in these terms, one can formally capture how 
coherent branching, adaptive control, and memory-dependent feedback 
collectively generate internally structured, higher-order dynamics. 
This offers a bridge between circuit-level models and abstract categorical 
semantics for quantum information \cite{Abramsky2004}.

Moreover, this perspective connects naturally to quantum-agent frameworks 
in which policy and memory registers guide adaptive decision-making 
\cite{Dunjko2016,Stromberg2024,Sultanow2025}. Viewed categorically, 
the QDM can be interpreted as an agent whose internal processes 
(coherently modulated by control, memory, and policy) instantiate a 
structured hierarchy of transformations, allowing formally defined 
deliberation and self-referential adaptation within a fully quantum substrate.

Taken together, these interpretations situate the QDM as more than 
just a technical construct. It is a unitary model of structured, adaptive 
dynamics, one that bridges concrete circuit realizations with abstract 
notions of higher-order quantum processes. It thus provides a simple 
but expressive model for analyzing self-modifying, deliberative 
dynamics within the language of quantum information theory.

\section{Conclusion}

We have proposed a toy model of a quantum deliberating machine, or 
QDM, in which internal deliberation is represented as a coherent superposition 
of competing unitary evolutions. The machine couples a system qubit to 
quantum control, memory, and policy registers, allowing branch histories 
to be tracked, adaptive strategies to evolve, and global coherence to be 
preserved. Individual memory registers appear classical in their reduced 
states, yet the full system retains entanglement, illustrating how quantum 
mechanics naturally encodes the history of internal decisions without violating 
the no-cloning theorem \cite{Wootters1982,Barnum1996}. Our explicit examples 
demonstrated that deterministic outcomes can arise in symmetric scenarios, 
while branch-dependent memory-based feedback generates probabilistic and 
entangled final states, highlighting the interplay between coherent deliberation, 
adaptive control, and feedback-modulated evolution. Our main contributions are: 
\begin{enumerate}
\item The formal definition of the QDM architecture $(C,M,S,P)$ for 
self-referential coherent feedback; 
\item The derivation of the ``Rescue'' protocol (Sec.\ \ref{sec:2-iter-example}), 
demonstrating how internal policy updates can steer a system to a target 
state regardless of initial branching.
\end{enumerate}

From an operational perspective, the QDM demonstrates that multiple 
alternative dynamical evolutions can coexist in superposition while still 
supporting branch-specific adaptation. Unlike classical parallel deliberation 
models, which rely on probabilistic sampling or repeated trials, the QDM 
simultaneously updates all alternatives coherently and reversibly. 
The policy register \(P\) adapts in response to the accumulated memory 
records, biasing future unitary evolutions of the system in a fully unitary 
and internal fashion. This, then, may be viewed as constituting 
\emph{adaptive deliberation}, rather than classical reinforcement, 
because no evaluative signal or reward is embedded in the update 
cycle, and the policy is not conditioned on external or measured 
outcomes. A generalization to \emph{reinforced deliberation} 
may also be implemented either internally by a slight modification 
of our main protocol (as described in Section \ref{sec:reinforcement} 
above), or by introducing explicit \emph{evaluative} feedback, internal 
or external, to bias future policies based on observed outcomes.

In addition, from the point of view of higher-order quantum maps, each 
iteration performed by the machine implements a process that acts on 
both states and prior processes, showing explicitly how 
operations can act on other operations \cite{Chiribella2008,Chiribella2009}. 
Categorical formulations may further illuminate the structured composition 
of these processes, highlighting the QDM’s place within a broader theoretical 
framework of quantum operations and process networks.

Future directions may include extending the model to longer sequences 
of iterations, richer or continuous feedback protocols, multi-qubit or 
continuous-variable systems, explicit connections to quantum learning 
and reinforcement learning formulations, and exploring physical 
implementations of the QDM model on near-term quantum processors, 
potentially using few-qubit systems to illustrate adaptive branching 
dynamics. The QDM provides a plausible setting for understanding how 
classical-like records and adaptive strategies may emerge from fully unitary, 
coherent quantum dynamics, offering a pathway for investigating internally 
parallel, reversible deliberation in quantum systems.

\appendix

\section{Controlled Stinespring unitaries in the pure-state QDM}
\label{sec:stinespring_pure_state}

In this Appendix we make explicit the connection between 
the QDM iteration map,
\[
  \mathcal{W}^{(k)} = \mathcal{V}^{(k)} \, 
  \mathcal{F}^{(k)} \, \mathrm{CNOT}_{C\to M^{(k)}} 
  \, \mathcal{U}^{(k)}.
\]
introduced in Section \ref{sec:summary-of-the-model} and 
the familiar Stinespring dilation picture of quantum channels. 
We present the argument entirely in the pure-state formalism, 
since in our work the QDM model is formulated solely in terms 
of global pure states and unitary operations on enlarged Hilbert 
spaces.

Let us first recall the Stinespring theorem \cite{Stinespring1955}: any 
Completely Positive Trace-Preserving (CPTP) map \cite{Kraus1971} 
on a system $S$ can be realized by a unitary operator on $S$ plus 
an auxiliary environment $E$ initialized in a fixed pure state, 
followed by discarding $E$. In symbols, for a channel 
$\mathcal{E}$ on $S$, there exists a unitary operator $U_{SE}$ 
and an environment pure state $\ket{0}_E$, such that for all density 
operators $\rho_S$,
\begin{equation}
  \mathcal{E}(\rho_S) = \mathrm{Tr}_E\!\big( U_{SE} \, 
  (\rho_S\otimes |0\rangle\langle 0|_E) \, U_{SE}^\dagger \big).
\end{equation}
In our pure-state QDM, we do not discard the auxiliary 
registers. Instead, they are retained as part of the global pure state,
which constitutes an important difference. The QDM's ``environment'' 
is internalized (memories and policy) and remains coherent across 
iterations.

More specifically, let $E$ denote an auxiliary Hilbert space that, 
for the QDM iteration $k$, we identify with the tensor product 
of the fresh memory register $M^{(k)}$ and the policy register 
$P$ (and possibly further ancillas). For each branch label 
$b$ in $\{0,1\}$ define a Stinespring unitary $U_b^{(k)}$ 
acting on $S\otimes E$ and an initial environment vector 
$\ket{0}_E$ (here $\ket{0}_E$ means $M^{(k)}$ in $\ket{0}$ 
and $P$ in its current value when appropriate). Then the 
controlled Stinespring unitary on $C\otimes S\otimes E$ is
\begin{equation}
  \mathcal{U}_{\rm ctrl}^{(k)} = \ket{0}\bra{0}_C 
  \otimes U_0^{(k)} + \ket{1}\bra{1}_C \otimes U_1^{(k)} .
\end{equation}
Acting on a product pure state 
$\big(\alpha\ket{0}_C+\beta\ket{1}_C\big)
\otimes\ket{\psi}_S\otimes\ket{0}_E$ this yields the global 
pure state,
\begin{equation}
  \mathcal{U}_{\rm ctrl}^{(k)}
   \Big( 
   \sum_{b=0}^1 \alpha_b \ket{b}_C \otimes \ket{\psi}_S \otimes \ket{0}_E 
   \Big)
  = 
  \sum_{b=0}^1 \alpha_b \ket{b}_C \otimes U_b^{(k)}
  \big( \ket{\psi}_S \otimes \ket{0}_E \big).
\end{equation}
If we now traced out $E$ we would recover the standard controlled 
channel on $S$; however, in the QDM construction we keep $E$ 
as part of the global state, so the state above remains pure and 
encodes both branch-dependent system evolution and the newly 
written memory/policy degrees of freedom.

The QDM iteration map $\mathcal{W}^{(k)}$ is realized by 
composing three kinds of controlled Stinespring unitaries 
(or simple controlled unitaries) all acting on the global pure 
state: 1) the controlled system evolution $\mathcal{U}^{(k)}$ 
(which is itself a controlled Stinespring unitary when the 
per-branch operations are dilations), 2) the coherent memory 
write $\mathrm{CNOT}_{C\to M^{(k)}}$ (which embeds 
branch information into $E$), and 3) the policy-controlled 
feedback $\mathcal{F}^{(k)}$ and policy update 
$\mathcal{V}^{(k)}$. Written as an overall unitary on the 
full Hilbert space (including previously written memory 
registers and the policy register), the iteration is simply 
a unitary operator,
\begin{equation}
  \mathcal{W}^{(k)} : 
  \mathcal{H}_{C}\otimes\mathcal{H}_{M^{(1)}\cdots M^{(k)}}
  \otimes\mathcal{H}_{S}\otimes\mathcal{H}_{P}
  \;\to\;
  \mathcal{H}_{C}\otimes\mathcal{H}_{M^{(1)}\cdots M^{(k)}}
  \otimes\mathcal{H}_{S}\otimes\mathcal{H}_{P},
\end{equation}
and it acts on global pure states. Because each building block 
is unitary, the composition is unitary and the global state 
remains normalized and pure at all times.

An alternative, but equivalent, viewpoint is to regard 
$\mathcal{W}^{(k)}$ as a controlled Stinespring dilation 
of an effective map on $S$, where the environment degrees 
of freedom are the internal registers $M^{(k)}$ and $P$. 
In that case the effective action on $S$ alone could 
be obtained (if desired) by taking the reduced state of $S$ 
via partial trace, but for the QDM we keep the full pure-state 
description, so that the machine's internal record and policy 
remain coherent resources for future iterations.

This controlled-Stinespring perspective clarifies two important 
conceptual points. First, it shows how the QDM remains fully 
compatible with standard quantum information theory; every 
step is a unitary on an enlarged Hilbert space, i.e., a Stinespring 
realization. Second, and most importantly for the QDM 
interpretation, it emphasizes that the ``environment'' in a QDM 
is internal and persistent rather than external and discarded. 
This persistence is what allows memory and policy to play 
an active, coherent, and self-referential role in subsequent 
iterations.

\section{Adaptive deliberation with outcome-driven 
oracle-based policy updates}
\label{sec:outcome-oracle-description}

In the canonical QDM model, the policy update at each 
iteration is based solely on the memory record of the action 
previously taken, not on the outcome of that action on 
the system state. This means it does not perform ``learning'' 
in the sense typically understood in machine learning or 
reinforcement learning: the policy is updated according to 
what was done, rather than what was achieved. The state of the 
system register---the actual outcome of the prior action---plays 
no role in modifying the policy.

To bring the QDM in line with the usual reinforcement-learning 
paradigm, one must properly close the feedback loop, so that 
the quality of an outcome (a reward signal) influences future 
policy. The goal here is to construct a minimal, fully unitary 
procedure capable of performing such \emph{outcome-driven 
policy updates}, while preserving quantum coherence and 
avoiding measurement collapse. The challenge is to have 
the system ``know'' its own outcome without performing an 
external measurement---that is, to enable learning without 
measuring.

A minimal and natural solution is to encode the reward signal 
as part of the joint quantum state itself, by adding two 
components to the canonical $(C,M,S,P)$ architecture:

\begin{enumerate}
\item A single-qubit \emph{reward register}, $R$, initialized to 
$\ket{0}_R$. It acts as a flag: if flipped to $\ket{1}_R$, it marks 
a ``successful'' or ``rewarded'' outcome within that branch of 
the superposition.

\item An \emph{outcome-evaluation oracle} 
$\mathcal{O}_{\rm rew}$, a unitary acting on the $SR$ subsystem 
that correlates $R$ with whether the system register $S$ 
is in the desired reward state $\ket{\psi_{\rm rew}}_S$. 
Formally,
\begin{equation}
\mathcal{O}_{\rm rew} = (I_S-P)\otimes I_R + P\otimes X_R, 
\qquad
P=\ket{\psi_{\rm rew}}\!\bra{\psi_{\rm rew}}_S,
\quad
\mathcal{O}_{\rm rew}^2=I,
\end{equation}
so that $R$ is flipped only when $S$ occupies the rewarded subspace. 
This is the standard ``flagging'' construction used throughout 
quantum algorithms---for example, in Grover-type oracles, 
where the marked (``good'') subspace acquires a phase or 
flag that can be coherently processed later.
\end{enumerate}

When the system is in a superposition of rewarded and 
non-rewarded states, $\mathcal{O}_{\rm rew}$ acts as a quantum 
sorter, entangling $S$ and $R$:
\begin{equation}
\ket{\Psi}_{SRP}
=
\left(a \ket{\psi_{\rm rew}}_S + b \ket{\psi_{\rm orthog}}_S\right)
\otimes \ket{0}_R \otimes \ket{\pi}_P
\; \longrightarrow \;
\mathcal{O}_{\rm rew}\ket{\Psi}_{SRP}
=
a \ket{\psi_{\rm rew}}_S \ket{1}_R \ket{\pi}_P
+
b \ket{\psi_{\rm orthog}}_S \ket{0}_R \ket{\pi}_P.
\end{equation}

The subsequent outcome-driven policy update is implemented 
as a controlled-unitary on $P$ with $R$ as the control,
\begin{equation}
  \tilde{\mathcal{V}}_{\rm adapt} 
  = \ket{0}\bra{0}_{R} \otimes \tilde{V}_0
  + \ket{1}\bra{1}_{R} \otimes \tilde{V}_1,
\end{equation}
giving
\begin{equation}
 \tilde{\mathcal{V}}_{\rm adapt}\mathcal{O}_{\rm rew}\ket{\Psi}_{SRP}
=
a \ket{\psi_{\rm rew}}_S \ket{1}_R \tilde{V}_1 \ket{\pi}_P
+
b \ket{\psi_{\rm orthog}}_S \ket{0}_R \tilde{V}_0 \ket{\pi}_P.
\end{equation}
Finally, the reward flag is uncomputed by reapplying the oracle,
\begin{equation}
 \mathcal{O}_{\rm rew} 
 \tilde{\mathcal{V}}_{\rm adapt}\mathcal{O}_{\rm rew}
 \ket{\Psi}_{SRP}
=
\left(
a \ket{\psi_{\rm rew}}_S \tilde{V}_1 \ket{\pi}_P
+
b \ket{\psi_{\rm orthog}}_S \tilde{V}_0 \ket{\pi}_P
\right)\otimes \ket{0}_R,
\end{equation}
so that $R$ is reset while the policy register has been coherently 
biased according to the reward structure.
Adopting $\ket{\psi_{\rm rew}}=\ket{1}_S$ yields the familiar 
CNOT-based flagging form:
\begin{equation}
\mathcal{O}_{\rm rew}
=
\ket{0}\bra{0}_S \otimes I_R 
+ \ket{1}\bra{1}_S \otimes X_R,
\end{equation}
exactly paralleling the oracle structure of standard 
quantum search or amplitude amplification procedures. 

\paragraph*{Connection to the two-iteration example.}
In the illustrative two-iteration model of the main text (Section \ref{sec:2-iter-example}), the 
rewarded state was chosen as $\ket{\psi_{\rm rew}}=\ket{1}_S$, 
but the control convention was reversed: the ``flag flip'' occurred 
for the \emph{non}-rewarded outcome, corresponding to an 
anti-CNOT version of $\mathcal{O}_{\rm rew}$. The two forms 
are unitarily equivalent up to relabeling the rewarded state 
($\ket{1}_S \leftrightarrow \ket{0}_S$) or by inserting $X_S$ gates 
before and after the oracle. Hence, both are physically equivalent 
and differ only by a choice of logical convention for ``success.''

\begin{tcolorbox}[float,floatplacement=t,colback=gray!5!white,colframe=gray!40!black,title={Equivalence Note: CNOT vs.\ anti-CNOT reward oracles}]
Two logically equivalent conventions can be used for defining the reward oracle $\mathcal{O}_{\rm rew}$:

\begin{enumerate}
\item[(i)] \textbf{CNOT form (standard reward flip):}
\[
\mathcal{O}_{\rm rew}
=
\ket{0}\bra{0}_S\otimes I_R
+
\ket{1}\bra{1}_S\otimes X_R,
\]
which flips the reward flag $R$ when $S=\ket{1}$, i.e.\ when the target is achieved.

\item[(ii)] \textbf{Anti-CNOT form (inverse convention):}
\[
\tilde{\mathcal{O}}_{\rm rew}
=
\ket{1}\bra{1}_S\otimes I_R
+
\ket{0}\bra{0}_S\otimes X_R,
\]
which flips $R$ when $S=\ket{0}$, i.e.\ when the target is \emph{not yet achieved}.
\end{enumerate}

The two are related by conjugation with $X_S$:
\[
\tilde{\mathcal{O}}_{\rm rew}
=
X_S\,\mathcal{O}_{\rm rew}\,X_S,
\]
and hence differ only by a relabeling of the rewarded basis state
($\ket{0}_S \leftrightarrow \ket{1}_S$).
All subsequent adaptive updates and interference effects remain identical up 
to this trivial change of convention.
For consistency, the canonical formulation in this paper adopts the
CNOT (``reward-when-$S=1$'') version, while the explicit two-iteration example 
employs the anti-CNOT variant for pedagogical symmetry.
\end{tcolorbox}

\paragraph*{Adaptive rotations of the policy register.}
To realize a small, continuous update of the policy, one may take
\begin{equation}
  \tilde{\mathcal{V}}_{\rm adapt} 
  = \ket{0}\bra{0}_{R} \otimes \tilde{R}_{y}(-\chi)_P
  + \ket{1}\bra{1}_{R} \otimes \tilde{R}_{y}(+\xi)_P,
\end{equation}
where $\tilde{R}_{y}(\pm\theta)$ denote opposite rotations on 
the Bloch sphere about the $y$-axis,
\begin{equation}
\tilde{R}_{y}(+\xi)_P
=
\begin{pmatrix}
\cos(\xi/2) & -\sin(\xi/2)\\
\sin(\xi/2) & \cos(\xi/2)
\end{pmatrix}_P,
\qquad
\tilde{R}_{y}(-\chi)_P
=
\begin{pmatrix}
\cos(\chi/2) & \sin(\chi/2)\\
-\sin(\chi/2) & \cos(\chi/2)
\end{pmatrix}_P.
\end{equation}
For small $\xi,\chi$, these rotations shift the policy amplitudes 
slightly toward the rewarded or non-rewarded poles of the 
Bloch sphere, thereby implementing a minimal, fully coherent 
form of outcome-based reinforcement.

Overall, this oracle-mediated enhancement enables the QDM 
to adapt its policy based on outcome quality, while maintaining 
global coherence and reversibility. The resulting single-iteration 
map becomes (cf.\ Eq.~\ref{fig:2nd-iter}),
\begin{equation}
  \label{eq:QDM_iteration_map_enhanced}
 \tilde{\mathcal{W}}^{(k)}
 =
 \mathcal{V}^{(k)} \,
 \left( \mathcal{O}^{(k)}_{\rm rew}
 \tilde{\mathcal{V}}^{(k)}
 \mathcal{O}_{\rm rew}^{(k)} \right) \,
 \mathcal{F}^{(k)} \, \mathrm{CNOT}_{C\to M^{(k)}}
 \, \mathcal{U}^{(k)},
\end{equation}
whose circuit diagram (for $\ket{\psi_{\rm rew}}=\ket{1}_S$) is 
shown in Fig.~\ref{fig:kth-iter-enhanced}.
\begin{equation}
  \label{eq:QDM_iteration_map_enhanced}
 \tilde{\mathcal{W}}^{(k)}
 =
 \mathcal{V}^{(k)} \,
 \left( \mathcal{O}^{(k)}_{\rm rew}
 \tilde{\mathcal{V}}^{(k)}
 \mathcal{O}_{\rm rew}^{(k)} \right) \,
 \mathcal{F}^{(k)} \, \mathrm{CNOT}_{C\to M^{(k)}}
 \, \mathcal{U}^{(k)},
\end{equation}
whose circuit diagram (with reward state chosen to be 
$\ket{\psi_{\rm rew}}=\ket{1}_S$) is given in 
Fig.\ \ref{fig:kth-iter-enhanced}.
\begin{figure}[H]
\centering
\scalebox{1}{\begin{quantikz}[row sep=0.5cm, column sep=0.6cm]
\lstick{$C$} & \ctrl{2} & \ctrl{1} & \qw& \qw& \qw & \qw & \qw \\
\lstick{$M^{(k)}$} & \qw & \targ{} & \qw& \qw& \qw & \ctrl{3} & \qw \\
\lstick{$S$} & \gate[wires=1]{U_0^{(k)}/U_1^{(k)}} & \gate[wires=1]{F_0^{(k)}/F_1^{(k)}} &\ctrl{1} & \qw& \ctrl{1} & \qw & \qw \\
\lstick{$R$} & \qw & \qw & \targ{} & \ctrl{1} & \targ{} & \qw & \qw \\
\lstick{$P$} & \qw & \ctrl{-2} & \qw& \gate[wires=1]{\tilde{V}_0^{(k)}/\tilde{V}_1^{(k)}} & \qw & \gate[wires=1]{V_0^{(k)}/V_1^{(k)}} & \qw
\end{quantikz}
}
\caption{Enhanced quantum circuit of the $k$-th QDM iteration 
incorporating the outcome-driven oracle-based policy update. 
With the reward state chosen as $\ket{\psi_{\rm rew}}=\ket{1}_S$, 
the oracle becomes the usual CNOT gate acting from $S$ to $R$.}
\label{fig:kth-iter-enhanced}
\end{figure}

\section{Summary of Cognitive Interpretation of the QDM Structure}

This Appendix maps the QDM’s structural components ($C, M, S, P$) to decision, memory, workspace, and policy functions familiar from computational and cognitive neuroscience. 

The explicit examples considered in the main text illustrate the operational layout and functioning of the Quantum Deliberating Machine (QDM). Beyond its formal definition in terms of registers and unitaries, it is useful to articulate how the distinct subsystems $C,M,S,P$ may be understood in cognitive terms. We propose the following interpretation:

\begin{itemize}
    \item[] \textbf{Control register $C$.} Formally, $C$ selects between two branches of evolution at each step, implementing a binary conditional. Cognitively, this register may be interpreted as a \emph{decision variable} or an \emph{attentional pointer}, representing the immediate choice or orientation of the system. It does not accumulate history by itself, but rather steers which pathway is followed.

    \item[] \textbf{Memory registers $M^{(k)}$.} In each iteration, the state of $C$ is copied into a fresh memory qubit. This creates a growing collection of entangled traces of prior control decisions. Formally, these registers preserve coherence of the branching structure. Cognitively, they are natural analogues of \emph{episodic or working memory}, i.e.\ explicit records of past choices and experiences. They do not remain passive, but rather influence later correlations and thereby embody the ``history dependence'' of cognition.

    \item[] \textbf{System register $S$.} This is the qubit on which all branch-dependent unitaries act, and which receives feedback from the policy register. Formally, it is the \emph{workspace} qubit whose reduced density matrix we track. Cognitively, it plays the role of the \emph{conscious workspace}, the foreground content of experience that is updated, reshaped, and continuously processed in time. Its trajectory is never autonomous, but always conditioned by both the current choice (via $C$) and the adaptive feedback (via $P$).

    \item[] \textbf{Policy register $P$.} Usually initialized independently of $C$, the policy is updated in every round through controlled unitaries conditioned on the system and control. Already from the first iteration onward, it exerts feedback on $S$. Formally, $P$ encodes a biasing unitary that is iteratively modified. Cognitively, $P$ is best interpreted as an \emph{internal model, belief state, or strategy representation}. It accumulates the system’s interaction history and, crucially, feeds back into future processing. This self-modifying character distinguishes the QDM from a simple branching decision tree.
\end{itemize}

Taken together, these correspondences motivate the following mapping:
\begin{align}
C &\;\mapsto\; \text{decision/attentional selector}, 
\nonumber \\
M & \;\mapsto\; \text{episodic traces}, 
\nonumber \\
S & \;\mapsto\; \text{conscious workspace}, 
\nonumber \\
P & \;\mapsto\; \text{adaptive strategy/belief model}.
\nonumber
\end{align}
In particular, the reflective channel of the policy register illustrates how the QDM can reinforce and refine its own strategies without invoking external update, exemplifying a self-referential mechanism by which adaptive structure is maintained within a purely unitary model. Correspondingly, the QDM is not merely a tree of conditional unitaries. The entanglement between control, memory, system, and policy implements both \emph{history} (via the $M^{(k)}$) and \emph{adaptivity} (via $P$). The system qubit $S$ therefore evolves under the joint influence of immediate choices and accumulated feedback. This renders the machine a toy model of self-referential deliberation, where memory, current state, and adaptive policy co-evolve rather than evolve in isolation.

\end{document}